\documentclass[prb,aps,twocolumn,floatfix,amsmath,amssymb]{revtex4-1}
\usepackage{latexsym}
\usepackage{graphicx}
\usepackage{tabularx}
\usepackage{times}
\usepackage[usenames,dvipsnames]{color}
\usepackage{amsmath}
\usepackage{dcolumn}
\usepackage{latexsym,amsmath,amssymb,bm,euscript}
\usepackage{comment}
\usepackage[
	colorlinks=true,
	urlcolor=BlueViolet,
    citecolor=Maroon,
	plainpages=false,
  	pdfpagelabels,
  	bookmarksnumbered
  	]{hyperref}

\newcommand{\ua}{\uparrow}
\newcommand{\da}{\downarrow}
\newcommand{\sgn}{\textrm{sgn}}

\renewcommand{\i}{\mathrm{i}}
\newcommand{\T}{\tilde{\mathcal{T}}}
\renewcommand{\d}{\partial}
\def\beq{\begin{equation}}
\def\eeq{\end{equation}}

\begin{document}

\title{Composite Dirac liquids: parent states for symmetric surface topological order}

\author{David F. Mross}
\thanks{These authors contributed equally to this work.}
\affiliation{Department of Physics and Institute for Quantum Information and Matter, California Institute of Technology,
Pasadena, CA 91125, USA}
\author{Andrew Essin}
\thanks{These authors contributed equally to this work.}
\affiliation{Department of Physics and Institute for Quantum Information and Matter, California Institute of Technology,
Pasadena, CA 91125, USA}
\author{Jason Alicea}
\affiliation{Department of Physics and Institute for Quantum Information and Matter, California Institute of Technology,
Pasadena, CA 91125, USA}

\begin{abstract}
{We introduce exotic gapless states---`composite Dirac liquids'---that can appear at a strongly interacting surface of a three-dimensional electronic topological insulator.  Composite Dirac liquids exhibit a gap to all charge excitations but nevertheless feature a single massless Dirac cone built from emergent \emph{electrically neutral} fermions.  These states thus comprise electrical insulators that, interestingly, retain thermal properties similar to those of the non-interacting topological insulator surface.  A variety of novel fully gapped phases naturally descend from composite Dirac liquids.  Most remarkably, we show that gapping the neutral fermions via Cooper pairing---which crucially does \emph{not} violate charge conservation---yields symmetric non-Abelian topologically ordered surface phases captured in several recent works.  Other (Abelian) topological orders emerge upon alternatively gapping the neutral Dirac cone with magnetism.  We establish a hierarchical relationship between these descendant phases and expose an appealing connection to paired states of composite Fermi liquids arising in the half-filled Landau level of two-dimensional electron gases.  To controllably access these states we exploit a quasi-1D deformation of the original electronic Dirac cone that enables us to \emph{analytically} address the fate of the strongly interacting surface.  The algorithm we develop applies quite broadly and further allows the construction of symmetric surface topological orders for recently introduced bosonic topological insulators.}
\end{abstract}
\maketitle

\section{Introduction}

Three-dimensional topological insulators (3D TIs)\cite{FuTI,MooreTI,RoyTI,KaneReview,QiReview} possess an electrically inert interior, yet harbor a wealth of physics at their surfaces that simply could not exist without the accompanying bulk.  For example, photoemission studies efficiently identify TI materials by detecting an odd number of massless surface Dirac cones\cite{Hsieh1,Hsieh2}---an impossible band structure in strictly two-dimensional, time-reversal-invariant media.  Introducing magnetism or superconductivity at the boundary gaps these Dirac cones but leaves nontrivial physics behind.  A magnetically gapped region of the surface enters an integer quantum Hall state with an anomalous \emph{half-integer} Hall conductance\cite{FuKaneSurfaceQHE}.  Domain walls between oppositely magnetized regions accordingly bind chiral gapless charge modes\cite{MajoranaModesWithChargeTransport,3DTI_Majorana_edge_states1}.  Gapping the surface instead by externally imposing Cooper pairing generates topological superconductivity in which vortices bind widely sought Majorana zero modes\cite{FuKane}.  The induced topological superconducting phase, while reminiscent of a spinless two-dimensional (2D) $p+ip$ superconductor\cite{ReadGreen}, preserves time-reversal symmetry and hence can not be `peeled away' from the TI bulk.  

With perturbatively weak electron-electron interactions the above discussion summarizes the essence of 3D TI boundary physics: massless Dirac cones imply preservation of both time reversal and charge conservation, while conversely a gapped surface necessitates breaking at least one of these symmetries.  But does symmetry imply surface metallicity more generally?  That is, with strong interactions is it possible to construct a \emph{symmetry-preserving, fully gapped} surface state for an electronic TI?  Several groups, quite remarkably, recently answered this question in the affirmative\cite{BondersonTO,WangTO,ChenTO,MetlitskiTO}, overturning conventional wisdom on the subject.  An important precursor to these studies originated from ingenious work on bosonic 3D topological insulators\cite{AshvinSenthil}, which were shown to support gapped surfaces that either break symmetries \emph{or} exhibit topological order that is forbidden in strictly 2D systems with the same symmetry.  Soon after symmetric 
gapped surface topological orders were identified for electronic TIs as well.  Interestingly, in contrast to the bosonic case these gapped surface states necessarily support non-Abelian anyons, emergent particles that enable inherently fault-tolerant storage and manipulation of quantum information\cite{kitaev,TQCreview}.  

\begin{figure*}
\centering
\includegraphics[width=6in]{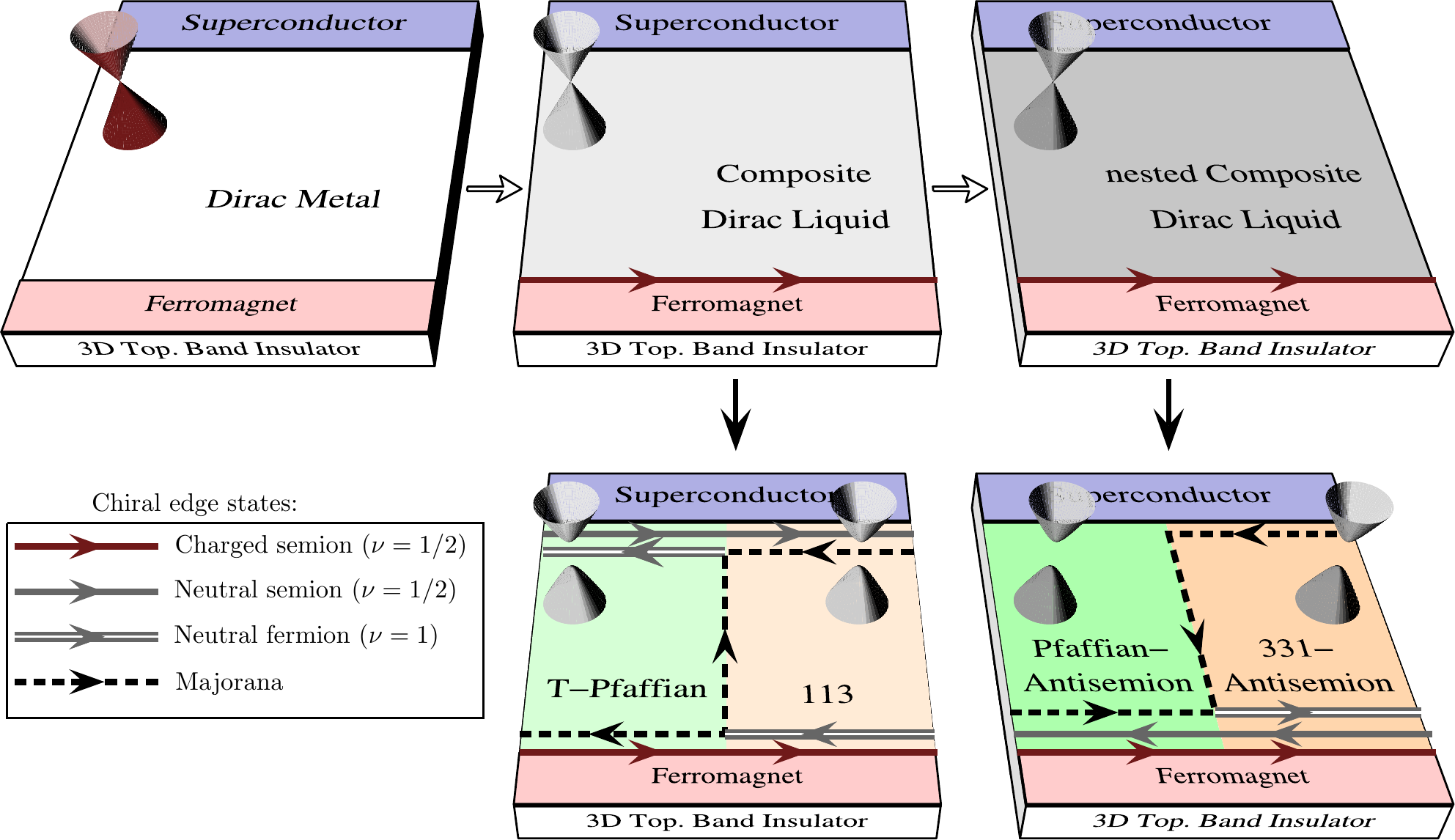}
\caption{Executive summary of results.  Stripping the electric charge off of the original surface Dirac cone (top left) yields a composite Dirac liquid (top center) that exhibits a charge gap but features a gapless Dirac cone formed by \emph{electrically neutral} fermions.  Analogously stripping a fictitious `pseudocharge' from the latter yields a nested composite Dirac liquid (top right) with a second-generation neutral Dirac cone.  These composite Dirac liquids serve as parent states for topologically ordered surface phases (bottom row) obtained by gapping the neutral Dirac cones with pairing or magnetism.  Importantly, pairing proceeds \emph{without} breaking electric charge conservation---because the paired fermions are neutral---and hence produces symmetric surface topological orders captured by previous works\cite{BondersonTO,WangTO,ChenTO,MetlitskiTO}.}
\label{SummaryFig}
\end{figure*}

Symmetric topological orders for the electronic TI surface have so far been accessed using two quite different means.  One approach starts from the broken-symmetry superconducting surface and then condenses a particular multiplet of vortices to restore charge conservation\cite{BondersonTO,WangTO,MetlitskiTO}; the other\cite{ChenTO} explicitly constructs the surface phases using so-called Walker-Wang models\cite{WalkerWang}.  The latter benefits from exact solvability but sacrifices direct connection to the original electronic degrees of freedom.  Two distinct, symmetric topological orders emerge from these approaches.  Using the language of Ref.~\onlinecite{ChenTO}, the phenomenological vortex condensation picture naturally accesses a `Pfaffian-antisemion' phase.  The Walker-Wang construction also captures this state but additionally reveals a `T-Pfaffian' phase with fewer quasiparticle types (see also Ref.~\onlinecite{BondersonTO}).  Both phases relate closely to the non-Abelian Moore-Read state\cite{
MooreRead}---also known as the Pfaffian---expected to occur at filling factor $\nu = 5/2$ in clean GaAs quantum wells.  We stress, however, that neither the T-Pfaffian nor the Pfaffian-antisemion topological order can appear in isolated 2D systems preserving time-reversal and charge-conservation symmetry.

In this paper we re-examine the strongly interacting TI surface using an alternative approach that both works directly with the original surface electrons \emph{and} follows a controlled, Hamiltonian-based formulation.  Our main result is the discovery of a new class of symmetric, strongly correlated gapless surface states that we christen composite Dirac liquids (CDLs).  Apart from exhibiting numerous striking physical response properties, these gapless states also constitute `parents' for both kinds of gapped symmetric topological orders noted above (among other nontrivial phases).  We will show that constructing CDLs allows us to capture the T-Pfaffian and Pfaffian-antisemion surface phases in a single, unified framework that, interestingly, reveals a hierarchical relationship between these topological orders.  

The simplest CDL follows by systematically stripping off the charge from the original surface Dirac electrons---leaving a \emph{neutral} massless Dirac cone immersed in a gapped, fractionalized, bosonic background that encodes the charge physics.  Because the charge sector acquires a gap the CDL is characterized by incompressible, electrically insulating behavior for the TI surface.  The neutral Dirac sea nevertheless yields a $T^2$ specific-heat contribution ($T$ denotes temperature) and underlies metallic longitudinal heat transport.  In other words, the CDL essentially retains the thermal but not charge characteristics of the original surface Dirac cone, thereby sharply violating the Wiedemann-Franz law.  

As our nomenclature suggests, the CDL reflects a surface analogue of the well studied composite Fermi liquid\cite{HLR} operative in the half-filled lowest Landau level for GaAs quantum wells, but (as usual) cannot live in 2D with the same symmetries.  Reviewing the latter provides a useful perspective on our results.  To understand the half-filled Landau level it proves exceedingly useful to decompose the electrons in terms of composite fermions bound to fictitious flux quanta that, on average, cancel the applied magnetic field.  Despite the strong Lorentz force acting on the electrons, the effectively neutral composite fermions move in straight lines over long distances and form a Fermi sea.  One can profitably view the composite fermion's Fermi sea as a 2D counterpart of the CDL's neutral Dirac cone.  (We stress however, that composite Fermi liquids---which are compressible, metallic states---exhibit rather different response properties, partly for symmetry reasons.)  Composite Fermi liquids serve as mother states for various interesting incompressible fractional quantum Hall phases.  Notably, the non-Abelian Moore-Read state arises upon gapping the Fermi sea by `weakly pairing' the composite fermions; forming a `strong pairing' condensate out of tightly bound composite-fermion pairs yields Abelian descendant quantum Hall states.\cite{ReadGreen}

Like its composite-Fermi-liquid-cousin, the CDL provides a convenient window for accessing nontrivial proximate phases for the electronic TI surface.  Most interesting, once we strip off the electric charge, the neutral Dirac cone can acquire a pairing gap \emph{without spoiling charge conservation symmetry}.  The resulting symmetric, gapped state is precisely the T-Pfaffian.  Conversely, we show that magnetically gapping the neutral Dirac cone instead yields a time-reversal-breaking Abelian topological order corresponding to the 113 state.  These CDL descendants are very similar in spirit to the weak- and strong-pairing phases\cite{ReadGreen} that derive from 2D composite Fermi liquids.  

Nesting the procedure above yields surface topological orders with richer structure.  In particular, one can form a new CDL out of the neutral Dirac cone rather than the original, charged Dirac electrons.  This nested CDL supports a second-generation neutral Dirac cone that coexists with a further fractionalized bosonic background.  Pairing the neutral fermions generates the Pfaffian-antisemion state captured by vortex condensation arguments, while magnetically gapping the second-generation neutral Dirac cone again yields an Abelian topological order.  There is, interestingly, a $Z_2$ aspect to the physics: nesting the CDL a second time yields adjacent gapped phases simply related to the T-Pfaffian and 113 states.  Figure \ref{SummaryFig} summarizes our main findings for CDLs and their descendant topological orders.

To formally establish these results we unapologetically abandon local electronic time-reversal symmetry $\mathcal{T}$ in our analysis.  Instead we enforce the weaker operation
\begin{equation}
  \tilde {\mathcal{T}} = \mathcal{T} \times T_y
  \label{T}
\end{equation}
corresponding to time reversal composed with a discrete translation $T_y$.  
Note that $\tilde {\mathcal{T}}$ symmetry protects the topological structure in certain `antiferromagnetic 3D topological insulators'\cite{AFTImong,AFTIessin,AFTIfang,spinlessAFTI}---relatives of the standard 3D TIs that host very similar physics including a quantized magnetoelectric effect.  Crucially, the $\tilde{\mathcal{T}}$-invariant surface states we construct \emph{also} can not appear in strictly 2D systems possessing the same symmetry.  
  
\begin{figure}
\centering
\includegraphics[width=\columnwidth]{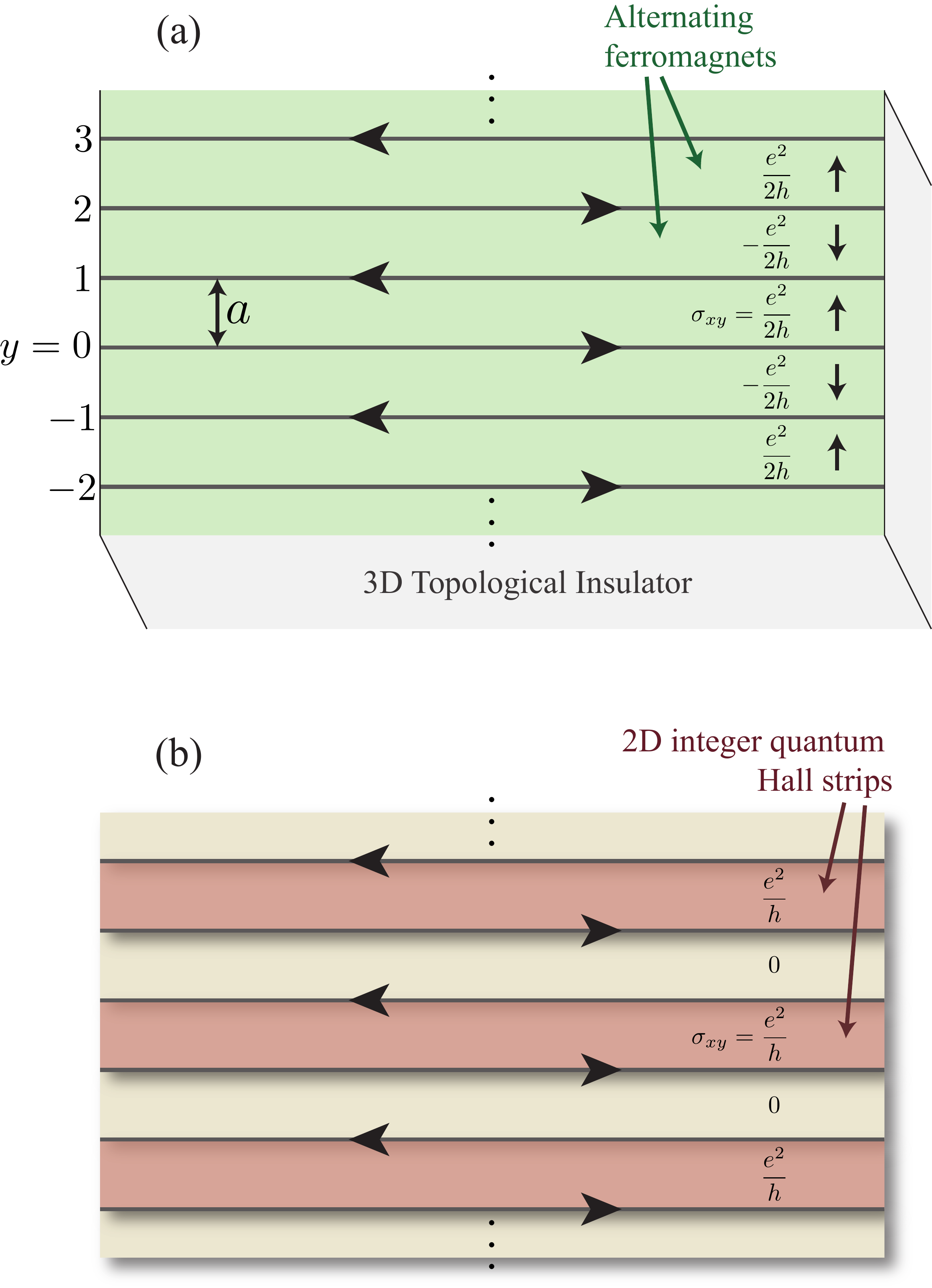}
\caption{(a) Strongly interacting coupled-chain setup analyzed in this paper.   Ferromagnetic strips of alternating magnetization reside on a 3D topological insulator surface.  Within each domain the electrons are gapped and exhibit a surface Hall conductance of $\sigma_{xy} = \pm e^2/(2h)$.  The domain walls, however, bind an array of gapless 1D electron modes with staggered chirality.  Importantly, the surface remains invariant under the symmetry $\T$ corresponding to time reversal composed with a translation by half a unit cell.  (b) The same set of gapless 1D modes arising in a strictly 2D setup consisting of $\nu = 1$ integer quantum Hall strips separated by insulators.  Contrary to the surface in (a), here $\T$ symmetry is always explicitly broken.  }
\label{Chains}
\end{figure}

Although we ultimately aim to capture nontrivial surface physics of ordinary (rather than antiferromagnetic) 3D TIs, relaxing $\mathcal{T}$ to $\tilde{\mathcal{T}}$ yields considerable technical advantages.  Most importantly, doing so allows us to map the problem onto weakly coupled `chains' and leverage the extensive theoretical machinery for one-dimensional systems to \emph{analytically} address the fate of the strongly interacting 2D surface.  Our method is inspired by pioneering studies that accessed quantum Hall phases from Luttinger liquid arrays\cite{KaneWires,TeoKaneChains}; see also Refs.~\onlinecite{LuWires,AshvinSenthil,QuantumWiresParafendleyons,Mong,Seroussi,Neupert,Sagi} for select related applications.  The mapping proceeds by overlaying the TI surface with ferromagnetic strips of equal width but alternating magnetization as shown in Fig.~\ref{Chains}(a).  As desired this setup preserves $\tilde{\mathcal{T}}$ with $T_y$ corresponding to translation by half a unit cell.  The domain walls 
bind charged fermion modes with staggered chirality that form our new low-energy degrees of freedom.  Incorporating $\tilde{\mathcal{T}}$-invariant electron tunneling between adjacent chiral modes simply restores the Dirac-cone band structure from which we began\cite{AshvinSenthil} (our setup indeed retains much of the TI surface physics).  Using bosonization, we can then controllably `neutralize' the Dirac cone, giving way to the CDLs and adjacent topologically ordered phases described above.  Properties of the ensuing gapped states follow very explicitly---including their quasiparticle content and the nature of gapless modes arising at boundaries with symmetry-breaking regions of the surface.  Similar constructions provide a recipe for accessing nontrivial surface physics in bosonic topological insulators\cite{AshvinSenthil}, and likely any setting where relaxing $\mathcal{T}$ to $\T$ does not coarsen the topological classification.

We structure the remainder of the paper as follows.  Section~\ref{Preliminaries} sets the stage by discussing some important details of electronic TI surface physics in the non-interacting limit.  We then develop the theory for the simplest composite Dirac liquid in Sec.~\ref{CDL} by following several complementary constructions that ultimately reproduce the same physics; Sec.~\ref{descendants} analyzes descendant phases of this CDL.  Nested CDLs and their descendants are briefly treated in Sec.~\ref{NestedCDL}.  All phases captured in this paper can be `peeled away' from the TI surface upon removal of $\T$ symmetry.  Section~\ref{2Dconnection} describes the mapping to pure 2D phases in this limit, which nicely sharpens the relationship to composite Fermi liquid physics alluded to above.  The general utility of our quasi-1D construction is established in Sec.~\ref{BosonicTO} where we employ a similar algorithm to recover symmetric surface topological orders for two different bosonic topological insulators.  Finally, we summarize our results and enumerate several open questions in Sec.~\ref{Discussion}.

\section{Preliminaries}
\label{Preliminaries}

Before diving into the CDL construction, we first review in greater depth the noninteracting TI surface properties summarized in the previous section.  
This discussion will lead us to the coupled-chain setup examined in later sections, and will also prove useful since much of the physics accessible from our CDLs can be readily understood by analogy with the non-interacting case.  
Consider a material such as Bi$_2$Se$_3$ whose boundary supports a single massless Dirac cone.\cite{Hsieh2}
The low-energy electron states at a surface in the $(x,y)$ plane are well described by a non-interacting Dirac Hamiltonian
\begin{align}
  H_{\rm surf} &= -i\hslash u \int_{x,y}\Psi^\dag (\partial_x \sigma^x + \partial_y \sigma^y) \Psi.
  \label{Hsurf}
\end{align}
Here $u$ is a velocity,  $\Psi(x,y)$ represents a two-component spinor comprised of spin up and down operators $\psi_\ua$ and $\psi_\da$, and $\sigma^{x,y,z}$ denote Pauli matrices that act in spin space.  
Equation~\eqref{Hsurf} preserves two important symmetries, U(1) charge conservation and (antilinear) time reversal $\mathcal{T}$, that act as follows:
\begin{align}
{\rm U(1)}: \Psi \rightarrow e^{i\alpha} \Psi, \quad \mathcal{T}: \Psi \rightarrow i \sigma^y \Psi 
\end{align}
with $\alpha$ an arbitrary constant.  Hence $\mathcal{T}^2$ sends $\Psi \rightarrow -\Psi$.  
Without adding interactions or breaking one of these symmetries, the surface spectrum necessarily remains gapless.  All mass-generating fermion bilinears indeed violate either U(1) or $\mathcal{T}$.  

It is worth elaborating on defects in the broken-symmetry phases as they play a central role throughout this paper. 
Adding an $s$-wave pairing term $\Delta \psi_\ua \psi_\da + \mathrm{H.c.}$ breaks U(1) and produces topological superconductivity at the surface.  A $2\pi$ vortex in the phase of the pairing potential $\Delta$ accordingly binds a Majorana zero mode that endows the vortices with projective non-Abelian statistics.  Gapping the surface via a Zeeman field $m \Psi^\dag \sigma^z \Psi$ yields a bulk magnetoelectric response described by axion electrodynamics with $\theta=\pi$,\cite{TopologicalFieldTheoryTI,EssinMooreVanderbilt} or alternatively by a surface quantum Hall response with Hall conductivity $\sigma_{xy} = {\rm sgn}(m)e^2/(2h)$.  The Hall conductivity jumps by $\pm e^2/h$ upon crossing oppositely magnetized regions with $m>0$ and $m<0$; consequently, their interface must host a chiral electron mode.  Assuming the domain wall resides along the $x$ direction the Hamiltonian for this chiral mode reads
\beq \label{eq:Hpsi}
H_{1d} = \int_x\, \psi^\dag (\mp i\hslash v_x \d_x) \psi,
\eeq
where $\psi$ describes a right- or left-moving electron with velocity $v_x$.  The unspecified sign in $H_{1d}$ dictates the chirality, which follows from the specific arrangement of Zeeman fields and reverses under $\mathcal{T}$.  Just as for the edge of a $\nu = 1$ integer quantum Hall state the domain wall is characterized by a quantized $e^2/h$ conductance.  
For later use we note that domain walls can bind multiple chiral electron modes in the more general case with an odd number of surface Dirac cones.

Consider next the situation where an \emph{antiferromagnetic} order parameter lifts time-reversal symmetry---possibly in the bulk as well as the surface.  One might naively expect the system to then smoothly connect to a trivial insulator.  This is certainly not the case, at least in the disorder-free limit.
Previous work has shown that the quantized magnetoelectric response remains protected by a residual symmetry given by time reversal followed by a lattice translation.\cite{AFTImong}  Consequently, surfaces that preserve this residual symmetry (along with charge conservation) still necessarily harbor a massless Dirac cone in the non-interacting limit.  This fact is exceedingly important since---purely as a theoretical crutch---we will only examine interacting surfaces that possess antiferromagnetic order in this paper.  Because the universal bulk and surface physics is nearly identical, our conclusions are expected to extend straightforwardly to TI surfaces respecting local time-reversal symmetry $\mathcal{T}$.  

We are most interested in surfaces where an imposed Zeeman field symmetrically alternates between $m > 0$ and $m < 0$ along the $y$ direction; see Fig.~\ref{Chains}(a).  In this case the relevant residual antiunitary symmetry is the modified time-reversal transformation $\tilde{\mathcal{T}}$ defined in Eq.~\eqref{T}.  While the electrons are gapped within each magnetized region, one can readily show that this configuration continues to host a single Dirac cone built from the chiral fermions bound to the domain walls.  Let $\psi_y$ be the fermion operator describing the chiral mode at domain wall $y$, indexed by integers as in Fig.~\ref{Chains}(a).  Even and odd $y$ respectively correspond to right- and left-movers.  Under $\tilde{\mathcal{T}}$ we thus have
\begin{equation}
  \T:  \psi_y \rightarrow (-1)^y \psi_{y+1}.
  \label{Tpsi}
\end{equation}
Note that $\T^2$ acting on real-space operators is not particularly illuminating since this operation reduces to a simple translation by one unit cell.  The action of $\T^2$ on momentum-space fields is, however, meaningful as we will see shortly.  

A minimal $\T$-invariant Hamiltonian that couples adjacent chiral modes is given by
\begin{align}
H &= \sum_y (-1)^y \int_x 
\left[ -i \hslash v_x \psi_y^\dagger \d_x \psi_y 
- t (\psi_y^\dag \psi_{y+1} + \mathrm{H.c.}) \right].
\label{Hchains}
\end{align}
The first term represents the kinetic energy while the second encodes nearest-neighbor hopping between domain walls with amplitude $t$.
Passing to momentum space yields
\begin{align}
\notag\\
H &= \sum_{k_y}\int_{k_x} \Psi^\dag_{\bf k}
\begin{pmatrix}
\hslash v_x k_x & -2 i t \sin{k_y a} \\
2 i t \sin{k_ya} & - \hslash v_x k_x
\end{pmatrix}
\Psi_{\bf k},
  \label{Hk}
\end{align}
where $a$ is the domain width and $\Psi_{\bf k} = (\psi_{e,k},\psi_{o,k})$ with $\psi_{e,k}$ ($\psi_{o,k}$) the Fourier transform of $\psi_y$ for $y$ even (odd); crucially, the ``unit cell'' has width $2a$, so $k_y$ takes values in $(-\pi/2a,\pi/2a]$.  As claimed, the Hamiltonian supports a single (anisotropic) Dirac cone centered at ${\bf k} = 0$, which one can see by retaining only small ${\bf k}$ modes in Eq.~\eqref{Hk}:
\begin{equation}
H \sim
\sum_{k_y}\int_{k_x}\Psi^\dag_{\bf k} \left( \hslash v_x k_x \sigma^z + \hslash v_y k_y \sigma^y \right) \Psi_{\bf k} .
  \label{HchainsDirac}
\end{equation}
On the right side $v_y = 2t a/\hslash$ denotes the velocity along $y$.  
At the Dirac point, $\T$ acts as
\begin{equation}
\T: \Psi_{(0,0)} \rightarrow i \sigma^y \Psi_{(0,0)} ,
\end{equation}
and so in the noninteracting, charge-conserving limit protects the gaplessness of $H$ by requiring the presence of a Kramers doublet.

Precisely the same Hamiltonian of Eq.~\eqref{Hchains} can describe a strictly 2D system composed of $\nu = 1$ integer quantum Hall strips (with $\sigma_{xy} = e^2/h$) bridged by ordinary insulators (with $\sigma_{xy} = 0$) as Fig.~\ref{Chains}(b) illustrates.  Contrary to our TI surface, however, $\T$ clearly \emph{never} exists as a microscopic symmetry for such a setup.  Viewed in the pure 2D context, Eq.~\eqref{Hchains} would be fine-tuned and describe the plateau transition between $\nu = 0$ and $\nu = 1$ integer quantum Hall states.  The construction above in fact essentially realizes the `network model' devised by Chalker and Coddington\cite{Chalker} to describe this very transition.  Vishwanath and Senthil took advantage of a similar correspondence to obtain a field theory for bosonic topological insulator surfaces in Ref.~\onlinecite{AshvinSenthil}.

The staggered-chirality TI surface modes described by Eq.~\eqref{Hchains} serve as our starting point for constructing exotic phases inaccessible from a weakly interacting perspective.  
Because we have deformed the original surface Hamiltonian such that it reduces to decoupled chains in a particular limit (i.e., $t = 0$) we can now bosonize to treat strong electron-electron interactions in a controlled manner.  In the following section we introduce a general prescription for `decorating' our staggered chiral fermions to neutralize the original surface Dirac cone, thereby entering a CDL state.  Descendant phases of the CDL are then analyzed in Sec.~\ref{descendants}.

\section{Composite Dirac liquid}
\label{CDL}

\begin{figure*}
\centering
\includegraphics[width=2\columnwidth]{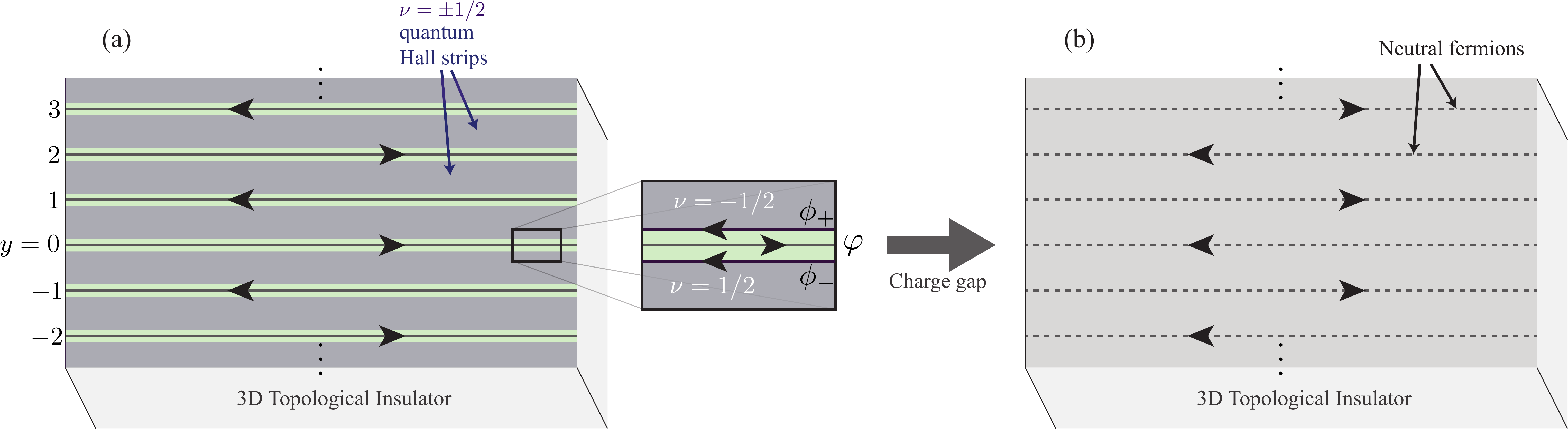}
\caption{(a) Magnetic domains of a TI surface overlaid with alternating $\nu = +1/2$ and $-1/2$ fractional quantum Hall fluids.  The quantum Hall strips cancel the half-integer Hall conductance from the magnetically gapped domains, so that $\sigma_{xy} = 0$ everywhere.  Each domain wall thus hosts a chiral electron native to the TI surface together with an opposing pair of quantum Hall charge modes (`spectator' neutral modes arising from the $\nu = \pm 1/2$ states are not shown for simplicity).  When interactions open a charge gap, a set of alternating-chirality \emph{neutral} fermions remains as in (b).  Tunneling between these modes that preserves $\T$ symmetry yields the neutral Dirac cone characteristic of the CDL state.  }
\label{ChargeGapFig}
\end{figure*}

We employ two complementary approaches to access a CDL state at the surface of a 3D TI.  The first, pursued in Sec.~\ref{extrinsic}, follows an `extrinsic' construction wherein we neutralize the Dirac cone by judiciously patterning the surface with 2D fractional quantum Hall states in a $\T$-invariant way.  See Fig.~\ref{ChargeGapFig}(a) for an example.  More precisely, we will show that hybridizing the chiral electron modes from our coupled-chain setup in Fig.~\ref{Chains}(a) with suitable quantum Hall edge states allows one to fully gap out the charge sector without breaking symmetries.  Very general considerations imply that a new set of low-energy \emph{neutral} fermionic modes with staggered chirality necessarily persist, however; $\T$-preserving hopping between these remnant modes yields precisely the CDL's neutral Dirac cone.  As we will see, this extrinsic construction benefits from a physically transparent, algorithmic line of attack.  Section~\ref{intrinsic} then develops a second, `intrinsic' 
approach.  There we capture the same CDL physics relying \emph{only} on interactions among chiral modes native to the TI surface (albeit in systems with three charged Dirac cones rather than one for technical reasons).

\subsection{Extrinsic construction}
\label{extrinsic}

\subsubsection{Dirac cone neutralization algorithm}
\label{algorithm}

The fact that each magnetically gapped domain in Fig.~\ref{Chains} breaks time reversal gives us enormous flexibility for decorating the surface to our convenience without further reducing its symmetries.  We are indeed free to arbitrarily deposit auxiliary charge-conserving systems into the `even' domains, so long as we insert their time-reversed counterparts in `odd' domains; such a setup generically preserves our modified time-reversal operation $\T$.  
Our aim here is to use these auxiliary degrees of freedom to facilitate gapping the charge sector.  With this in mind, recall that chiral electron modes necessarily bind to the domain walls in Fig.~\ref{Chains}(a) because the surface Hall conductivity jumps by $\pm e^2/h$ upon crossing oppositely magnetized domains. 
Suppose then that we tile the domains with quantum Hall fluids whose filling alternates between $\nu = +1/2$ and $-1/2$, such that the overlaid quantum Hall states exactly cancel the half-integer Hall conductance from the magnetically gapped regions of the TI surface.  Since $\sigma_{xy} = 0$ everywhere, domain walls need no longer support gapless charge modes.  Upon closer inspection we can show that the desired charge gap can always arise from physical electronic processes in configurations exploiting any Abelian $\nu = \pm 1/2$ quantum Hall phases (and likely any non-Abelian ones as well).

As Fig.~\ref{ChargeGapFig}(a) illustrates, each domain wall now hosts a chiral electron mode from the original TI surface together with \emph{two} sets of quantum Hall edge states---one from each adjacent domain.  Together, these gapless modes contribute a charge conductance that vanishes as required for opening a charge gap without breaking U(1) symmetry.  
Let us temporarily focus on a single domain wall, say $y = 0$.  
From hydrodynamic arguments due to Wen,\cite{WenBook} we know that the chiral charge modes at the upper and lower quantum Hall edges of the $y = 0$ domain wall can be captured by chiral boson fields $\phi_+$ and $\phi_-$, respectively [see Fig.~\ref{ChargeGapFig}(a)].  These fields describe left-moving `edge semions' and obey the commutation relation (Kac-Moody algebra)   
\beq
[\partial_x \phi_\pm (x), \phi_\pm (x')] = -i \pi \delta(x-x').
\eeq
For electronic $\nu = \pm 1/2$ states neutral modes (which we leave unspecified for now) generically coexist with these charge modes.  
We can, moreover, bosonize the right-moving chiral electron $\psi$ at the TI magnetic domain wall by writing $\psi \sim e^{i \varphi}$, where $\varphi$ satisfies
\begin{equation}
  [\partial_x\varphi(x),\varphi(x')] = i 2 \pi \delta(x-x').
  \label{phicommutator}
\end{equation}

In terms of the above fields the electron density becomes $\rho = \partial_x\Theta_c/\pi$ with
\begin{align}
  \Theta_c = (\varphi - \phi_+ - \phi_-)/2 
  \label{Theta}
\end{align}
a non-chiral field whose dual is given by
\begin{equation}
  \Phi_c = (\varphi + \phi_+ + \phi_-)/2.
\end{equation}
Note that charge conservation symmetry shifts $\Phi_c$ by a constant but leaves $\Theta_c$ invariant.  Thus \emph{if} physically permissible interactions can pin $\Theta_c$ then the charge sector will acquire a gap while preserving U(1).
It remains to determine whether such terms actually appear in the Hamiltonian.  This is not \emph{a priori} obvious since only electrons (and not fractional excitations) can hop between the quantum Hall edges and the magnetic domain wall, but turns out to be the case quite generally.  

Let $K$ and $\vec q$ respectively denote the K-matrix and charge vector characterizing the (Abelian) quantum Hall state just below the domain wall.  The full set of edge fields $\vec \phi$ for this state---including now neutral modes---exhibits commutation relations $[\partial_x\phi_I(x),\phi_J(x')] = -i2 \pi K^{-1}_{IJ}\delta(x-x')$.  Physical quasiparticle operators take the form $e^{i\vec l \cdot \vec \phi}$ with $\vec l$ an arbitrary integer vector.  Such operators add charge $\vec l \cdot K^{-1}\cdot \vec q$ to the edge.  In this language electrons are created by $e^{i \vec m \cdot K \cdot \vec \phi}$, with $\vec m$ another integer vector satisfying $\vec m\cdot \vec q = 1$, while the charge field defined earlier reads $\phi_- = \vec q \cdot \vec \phi$.  We make two noteworthy observations: $(i)$ $\phi_-$ trivially commutes with all neutral operators so that the charge and neutral sectors can in principle be gapped independently and $(ii)$ there always exists an integer $n$ such that $e^{i n \phi_-}$ can be 
expressed as a product of electron operators.  The same discussion of course applies straightforwardly to the quantum Hall edge just above the domain wall.  

Point $(ii)$ above together with Eq.~\eqref{Theta} imply that for some integer $n$ a term
\begin{equation}
  H_{\rm charge~gap} = -\lambda \int_x \cos(2n\Theta_c)
  \label{charge_gap}
\end{equation}
arises from physically allowed, correlated hopping processes that symmetrically transfer $n$ electrons from the magnetic domain wall into the adjacent quantum Hall edges.  Sufficiently strong repulsive interactions (which we invoke hereafter) can always make this term relevant, in which case $\Theta_c$ locks to the minima of Eq.~\eqref{charge_gap} and the charge sector gaps out.  Massive charge-$e/n$ excitations create $\pi/n$ kinks in $\Theta_c$ that interpolate between adjacent minima of the cosine potential.  
At this point the auxiliary quantum Hall systems have dutifully fulfilled their purpose.  We assume that all remaining neutral modes that they carry subsequently also gap out via tunneling of neutral quasiparticles across each quantum Hall strip.\footnote{For consistency it is important that the charge sector gaps out at higher energy scales than the neutral sector, since otherwise quasiparticle tunneling across the quantum Hall strips would compete with the charge-gapping $\lambda$ term, possibly yielding different physics.  Such a hierarchy is reasonable since $\lambda$ arises from purely local electronic processes within a domain wall.} 

Crucially, we have not fully gapped out the $y = 0$ domain wall.  Recall that we gapped the overall charge mode by hybridizing a \emph{pair} of left-moving quantum Hall charge modes with just \emph{one} right-moving electron from the magnetic domain wall.  In total these modes yield a vanishing charge conductance but a nonzero quantized thermal Hall conductance\cite{KaneThermal} (or equivalently chiral central charge).  Thus a single left-moving gapless neutral mode necessarily persists.  This neutral mode is described by a field $\tilde\varphi = \phi_+-\phi_-$ with commutation relation
\begin{equation}
  [\partial_x\tilde \varphi(x),\tilde \varphi(x')] = - i2\pi \delta (x-x'),
\end{equation}
identical to Eq.~\eqref{phicommutator} but with opposite sign.  \emph{The original TI surface electron $\psi \sim e^{i \varphi}$ has thus been effectively converted into a neutral fermion $\tilde \psi \sim e^{i \tilde \varphi}$ with reversed chirality.}  One can view $\tilde \psi$ as an electric dipole formed by semions from the adjacent quantum Hall edges.  

Because the properties of all other domain walls follow trivially by $\T$ symmetry, we can immediately paint a more complete picture for the surface's residual low-energy physics.  The domain walls now host an array of neutral fermions $\tilde \psi_y$ with alternating chirality, very similar to the set of chiral electron modes from which we started; compare Figs.~\ref{Chains}(a) and \ref{ChargeGapFig}(b).  \emph{Within the ground state for the charge sector}, hopping among the neutral fermions accordingly can be described by a straightforward modification of Eq.~\eqref{Hchains}:
\begin{align}
H_{\rm neut} &= \sum_y (-1)^y \int_x 
\left[ +i \hslash \tilde v_x \tilde \psi_y^\dagger \d_x \tilde \psi_y 
- \tilde t (\tilde \psi_y^\dag \tilde \psi_{y+1} + \mathrm{H.c.}) \right].
\label{Hchains2}
\end{align}
Here $\tilde v_x$ and $\tilde t$ respectively represent the neutral fermion velocity along $x$ and nearest-neighbor tunneling strength.  The Hamitonian above supports a `neutralized' Dirac cone centered around zero momentum; this can be easily verified by following the steps leading from Eq.~\eqref{Hchains} to \eqref{HchainsDirac}.  We have thus successfully stripped off the charge from the original Dirac cone and entered the CDL we sought to construct.  

We stress that the neutral fermions do \emph{not} completely decouple from the charge sector.  As we will see explicitly below, a careful derivation of the neutral fermion hopping Hamiltonian shows that charge excitations---neglected when writing Eq.~\eqref{Hchains2}---appear as flux seen by the neutral fermions.  This fact will prove essential when studying properties of the topologically ordered descendant phases of the CDL.

\subsubsection{331-based setup}

To put the preceding discussion on firmer footing, we now apply the general algorithm outlined above to a setup in which the $\nu = \pm 1/2$ quantum Hall fluids realize Halperin 331 states (and their time-reversed conjugates).  
The 331 state ranks as one of the simplest Abelian $\nu = 1/2$ electronic quantum Hall phases and is characterized by a two-dimensional K-matrix
\begin{align}
 K = \begin{pmatrix} 3 & 1 \\ 1 & 3\end{pmatrix}
\end{align}
and charge vector $\vec q = (1,1)$.  
We denote the fields for the upper $(+)$ and lower $(-)$ quantum Hall edges at domain wall $y$ by $\vec \phi_{\pm,y} = (\phi_{1\pm,y},\phi_{2\pm,y})$.  
These fields satisfy nontrivial commutation relations
\begin{equation}
  [\partial_x\phi_{Is,y}(x),\phi_{Js,y}(x')] = -i 2\pi (-1)^y K^{-1}_{IJ}\delta(x-x')
\end{equation}
and can be used to construct elementary electron operators via
\begin{align}
  \psi_{1s,y} = e^{i (3 \phi_{1s,y} + \phi_{2s,y})} \nonumber
  \\
  \psi_{2s,y} = e^{i (\phi_{1s,y} + 3\phi_{2s,y})}
\end{align}
with $s = \pm$.  
In addition to the edge fields the magnetic domain walls support chiral electrons $\psi_y \sim e^{i\varphi_y}$, where
\begin{equation}
  [\partial_x\varphi_y(x),\varphi_y(x')] = i2 \pi(-1)^y \delta(x-x').
  \label{phicommutator2}
\end{equation}
The factors of $(-1)^y$ in the commutators reflect the alternating chirality of the modes along $y$.  

We will find it very convenient to repackage the fields as follows.  As in Sec.~\ref{algorithm} we define non-chiral charge-sector fields
\begin{eqnarray}
  \Theta_{c,y} &=& \left[\varphi_y - \vec q \cdot \left(\vec \phi_{+,y} + \vec \phi_{-,y}\right)\right]/2
  \nonumber \\
  \Phi_{c,y} &=& \left[\varphi_y + \vec q \cdot \left(\vec \phi_{+,y} + \vec \phi_{-,y}\right)\right]/2;
  \label{ThetaPhi}
\end{eqnarray}
the total electron density at domain wall $y$ then reads $\rho_y = (-1)^y \partial_x \Theta_{c,y}/\pi$. 
The remaining neutral sector is conveniently decomposed in terms of (important) neutral fermions
\begin{equation}
  \tilde \psi_y \sim e^{i \tilde \varphi_y} \equiv e^{i \vec q\cdot\left(\vec \phi_{+,y} - \vec \phi_{-,y}\right)}
\end{equation}
and (largely spectator) non-chiral `flavor' fields
\begin{eqnarray}
  \Theta_{f,y+\frac{1}{2}}&=& [(\phi_{1+,y}- \phi_{2+,y})- (\phi_{1-,y+1}-  \phi_{2-,y+1})]/2
  \nonumber \\
  \Phi_{f,y+\frac{1}{2}} &=& [(\phi_{1+,y}- \phi_{2+,y})+ (\phi_{1-,y+1}-  \phi_{2-,y+1})]/2.
  \nonumber \\
  \label{FlavorFields}
\end{eqnarray}
Notice that the flavor degrees of freedom involve quantities from domain walls $y$ and $y+1$; hence we index them by half-integers $y+1/2$.  The charge fields commute with all neutral fields.\footnote{This is true modulo Klein factors, which we do not keep track of here because they do not affect our results.}  

Let us express the full $\T$- and U(1)-invariant Hamiltonian that puts the surface into a CDL state as 
\begin{equation}
  H = H_0 + H_c + H_f + H_n.  
\end{equation}
The first term, $H_0$, contains quadratic couplings for the bosonized fields encoding various density-density interactions and will not be written explicitly.  Correlated hoppings that open a charge gap appear in the second term,
\begin{eqnarray}
  H_c &=& \sum_y \int_x \left(\lambda \psi^{4}_y\psi_{1+,y}^\dagger\psi_{2+,y}^\dagger \psi_{1-,y}^\dagger \psi_{2-,y}^\dagger+h.c.\right)
  \nonumber \\
  &\sim& -\lambda \sum_y\int_x \cos(8\Theta_{c,y})
  \label{Hc}
\end{eqnarray} 
(Again we implicitly assume strong repulsive density-density interactions in $H_0$ so that the $\lambda$ term is relevant.  Moreover, quantities like $\psi^4$ should be interpreted as appropriately regularized.)
The third term gaps the neutral modes native to the 331 states,
\begin{equation}
  H_f = -\kappa \sum_y \int_x\cos\left(2\Theta_{f,y+\frac{1}{2}}\right);
\end{equation}
this too is presumed relevant.
From Eqs.~\eqref{FlavorFields} we see that $H_f$ arises from tunneling of fractional quasiparticles from one edge of a given quantum Hall strip to the other.  These processes are physical since the intervening fluid is fractionalized.  Once gapped the flavor degrees of freedom play absolutely no role in either the CDL or the descendant topological orders.  For simplicity we thus henceforth take without loss of content
\begin{equation}
  \Theta_{f,y+\frac{1}{2}} = 0.
  \label{Thetaf}
\end{equation}
Finally, $H_n$ encodes the kinetic energy for the neutral fermions $\tilde \psi_y$.  Some care is needed to ensure that this piece reflects physical electronic processes between domain walls.

Interchain neutral fermion tunneling arises from correlated electron hopping events of the following type:
\begin{equation}
  (\psi_{1+,y}^\dagger \psi^\dagger_{2-,y+1})(\psi_y \psi_{y+1}) \sim e^{2i(\Theta_{c,y}+\Theta_{c,y+1})}\tilde \psi_y^\dagger \tilde \psi_{y+1}
  \label{NeutralFermionHopping}
\end{equation}
[When bosonizing we used Eq.~\eqref{Thetaf}.]  As remarked at the end of Sec.~\ref{algorithm} we indeed see that the neutral fermions couple to the charge sector through the phase factor on the right side.  The neutral fermion kinetic energy thus reads
\begin{eqnarray}
  H_n &=& \sum_y (-1)^y \int_x \bigg{\{} +i \hslash \tilde v_x \tilde \psi_y^\dagger \d_x \tilde \psi_y 
  \nonumber \\
  &-& \tilde t \left[\tilde \psi_y^\dag \tilde \psi_{y+1}e^{2i(\Theta_{c,y}+\Theta_{c,y+1})} + \mathrm{H.c.}\right] \bigg{\}}.
\label{Hchains3}
\end{eqnarray}
Equations~\eqref{Hc} and \eqref{Hchains3} correctly capture the physics of the CDL---including low-lying charge excitations.

\subsubsection{Moore-Read-based setup}

In the previous discussion, we obtained a neutral Dirac cone by essentially stripping the charge off of the magnetic domain wall electrons with the help of \emph{Abelian} $\nu = \pm 1/2$ edge states.  
It is natural to ask whether alternatively employing \emph{non-Abelian} edges yields a surface theory with richer structure.  At least for the simplest example where the surface is tiled with Moore-Read states (and the time-reversed conjugate) this is certainly not the case.  Here the edge theory consists of a $\nu=1/2$ charged boson and a copropagating Majorana mode.  Just as in the Abelian 331-based construction, a charge gap opens by tunneling four electrons from a given magnetic domain wall symmetrically into the two adjacent Moore-Read edges.  Also as in the 331 case the remaining neutral modes---in this case Majoranas---are spectators that can be trivially gapped by tunneling across the domains.  The non-Abelian excitations originating from the Moore-Read states confine and do not appear in the low-energy surface theory.  What remains is a CDL, again described by Eqs.~\eqref{Hc} and \eqref{Hchains3}.

Experts will also note that a particularly clever choice of edge theory would allow us to gap the domain walls immediately.  This edge theory is that of the `T-Pfaffian,'\cite{BondersonTO,ChenTO} to which we will return in Sec. \ref{descendants}.  However, for our purposes this is actually undesirable, since we would miss the novel CDL state entirely, and the ability to access the variety of descendant phases that we discuss in later sections.

\subsection{Intrinsic construction}
\label{intrinsic}

Largely for theoretical expedience we so far captured the CDL by adding fractionalization into the TI surface by hand.  This is by no means necessary.  Here we develop an `intrinsic' construction that arrives at the same physics starting from \emph{unfractionalized} electronic degrees of freedom.  Consider a TI material whose surface supports three charged Dirac cones (some remarks on the single-cone case appear below).  Each magnetic domain wall from Fig.~\ref{Chains}(a) correspondingly now carries three copropagating chiral electrons that we describe with operators
 \begin{equation}
  \psi_y \sim e^{i\varphi_y}, \quad \psi_{\pm,y} \sim e^{i\phi_{\pm,y}}.
  \label{IntrinsicElectrons}
\end{equation}
All three bosonized fields satisfy commutation relations identical to Eq.~\eqref{phicommutator2}.  
It is useful to imagine the $\psi_{+,y}$ and $\psi_{-,y}$ electron modes as residing just above and below a central $\psi_y$ mode.  
One can view the former as auxiliary fields that play a similar role in the theory as the fractional quantum Hall edges in our `extrinsic' constructions.  

The key step forward is finding a suitable basis change to nonchiral fields that can develop a gap while preserving $\T$ and U(1) symmetries, along with a neutral fermion that remains gapless.  Such a basis is less obvious here due to the necessity of mixing fields from multiple domain walls.  As a hint, notice that the combinations $(3\phi_{-,y+1} - \phi_{+,y})/4$ and $(3\phi_{+,y-1} - \phi_{-,y})/4$ exhibit `semionic' commutation relations for odd $y$ and `antisemionic' for even $y$---exactly as for the charge fields from the quantum Hall edges examined earlier.  This correspondence suggests defining non-chiral `charge sector' fields via
\begin{eqnarray}
  \Theta_{c,y} &=& [\varphi_y-(3\phi_{-,y+1} - \phi_{+,y} + 3\phi_{+,y-1} - \phi_{-,y})/4]/2
  \nonumber \\
  \Phi_{c,y} &=& [\varphi_y+(3\phi_{-,y+1} - \phi_{+,y} + 3\phi_{+,y-1} - \phi_{-,y})/4]/2,
  \nonumber \\
\end{eqnarray}
directly parallel to Eqs.~\eqref{ThetaPhi}.  Indeed $\Theta_{c,y}$ and $\Phi_{c,y}$ defined as above exhibit precisely the same commutation relations as before (though of course the electron density is no longer proportional to $\partial_x \Theta_{c,y}$).  Moreover, both commute with the linearly independent field
\begin{equation}
  \tilde \varphi_y = (3\phi_{-,y+1} - \phi_{+,y} - 3\phi_{+,y-1} + \phi_{-,y})/4
  \label{varphiIntrinsic}
\end{equation}
that defines a neutral fermion through $\tilde \psi_y \sim e^{i \tilde \varphi_y}$.  

We simply need to verify that the Hamiltonian terms $H_c$ and $H_n$ [Eqs.~\eqref{Hc} and \eqref{Hchains3}] that put the surface into the CDL arise from physical electronic processes in this realization.  The `charge sector' acquires a gap through the multi-electron term
\begin{equation}
  \psi_{+,y-1}^{\dagger 3} \psi_{-,y} \psi_y^{4} \psi_{+,y} \psi_{-,y+1}^{\dagger 3} + \mathrm{H.c.} \sim \cos(8\Theta_{c,y}).
  \label{ChargeGapIntrinsic}
\end{equation}
Furthermore the neutral fermion's kinetic energy along $y$ in Eq.~\eqref{Hchains3} arises from correlated electron hoppings
\begin{equation}
  (\psi_{+,y}^\dagger \psi_{-,y+1}^\dagger)(\psi_y  \psi_{y+1})  \sim e^{2i(\Theta_{c,y} + \Theta_{c,y+1}) } \tilde \psi^\dagger_y \tilde \psi_{y+1}\label{NeutralFermionHoppingIntrinsic};
\end{equation}
note the similarity to Eq.~\eqref{NeutralFermionHopping}.  This demonstrates that the CDL is indeed accessible from our intrinsic construction.  

What about surfaces like in Bi$_2$Se$_3$ that host just a single Dirac cone?  While the CDL is undoubtedly accessible here too, our quasi-1D approach does not provide a convenient entrance (at least without reducing translation symmetry).  The reason why is clear: there are simply too few fields per domain wall.  Consequently, one can not independently define a `charge sector' that gaps out, leaving an independent neutral fermion that one can then liberate between chains.  Rather, interactions must, in a sense, induce both effects simultaneously.  Overcoming this problem would be very interesting but will not be pursued further in this paper.

\subsection{Universal properties of the composite Dirac liquid}
\label{sec:universalcdl}

Within any of our constructions the universal physics of the CDL can be described by a Hamiltonian 
\begin{eqnarray}
  H_{\rm CDL} &=& H_c + H_n
  \label{HCDL} \\
  H_c &=& -\lambda \sum_y\int_x \cos(8\Theta_{c,y})
  \\
  H_n &=& \sum_y (-1)^y \int_x \bigg{\{} +i \hslash \tilde v_x \tilde \psi_y^\dagger \d_x \tilde \psi_y 
  \nonumber \\
  &-& \tilde t \left[\tilde \psi_y^\dag \tilde \psi_{y+1}e^{2i(\Theta_{c,y}+\Theta_{c,y+1})} + \mathrm{H.c.}\right] \bigg{\}},
\end{eqnarray}
where again $\tilde \psi_y\sim e^{i\tilde \varphi_y}$ represents a neutral fermion and $\Theta_{c,y}$ is a charge field specifying the electron density via $\rho_y = (-1)^y \partial_x \Theta_{c,y}/\pi$.  Observe that for simplicity we have suppressed all charge-sector dynamics so that $\Theta_{c,y}$ rigidly locks to $\pi/4$ times an integer.\footnote{Technically we should include Hamiltonian terms involving $(\partial_x\Theta_{c,y})^2$ and $(\partial_x\Phi_{c,y})^2$ so that kinks in $\Theta_{c,y}$ cost energy.  Such terms can be trivially added if desired.}  The dual charge field $\Phi_{c,y}$ increments the electron density by virtue of the commutator
\begin{equation}
  [\partial_x\Phi_{c,y}(x),\Theta_{c,y}(x')] = i \pi (-1)^y \delta(x-x')
\end{equation}
That is, the operator $e^{i q \Phi_{c,y}}$ adds $q$ units of electric charge to domain wall $y$.  

One can revealingly recast the problem in terms of a `dressed' neutral fermion
\begin{equation}
  \tilde \Psi_y \equiv \tilde \psi_y e^{2i \sum_{y' < y}(\Theta_{c,y'}+\Theta_{c,y'+1})} \label{eqn:dressedfermion}.
\end{equation}
The $\Theta_{c,y}$ string introduced above cancels the phase factors in the interchain tunneling term, yielding
\begin{eqnarray}
  H_n = \sum_y (-1)^y \int_x \left[ +i \hslash \tilde v_x \tilde \Psi_y^\dagger \d_x \tilde \Psi_y 
  - \tilde t \left(\tilde \Psi_y^\dag \tilde \Psi_{y+1}+ \mathrm{H.c.}\right) \right].
\label{Hchains4}
\end{eqnarray}
In this form one can readily take the continuum limit along $y$ by writing
\begin{equation}
  \tilde \Psi_y(x) \sim\begin{cases}
    \tilde \Psi_\uparrow({\bf r}),& \text{for}~y~\text {even}\\
    \tilde \Psi_\downarrow({\bf r}),& \text{for}~y~\text {odd}.
\end{cases}
\end{equation}
Equation \eqref{Hchains4} then explicitly takes the form of a single neutral Dirac cone:
\begin{eqnarray}
  H_n \sim \int_{x,y} \tilde\Psi^\dagger(i\hslash \tilde v_x \partial_x\sigma^z -i\hslash \tilde v_y \partial_y \sigma^y)\tilde \Psi,
\label{Hchains5}
\end{eqnarray}
where $\tilde v_y \propto \tilde t$.  We emphasize that despite appearances the neutral and charge sectors remain linked, as charge excitations alter the boundary conditions on $\tilde \Psi_y$.  In particular, a dressed neutral fermion moving around a localized $e/4$ excitations---corresponding to a $\pi/4$ kink in $\Theta_{c,y}$---acquires a minus sign.  

The CDL Hamiltonian exhibits three important symmetries: U(1) electric charge conservation, $\T$, and a symmetry U(1)$_n$ corresponding to neutral-fermion conservation.  The first of these is preserved trivially since U(1) modifies neither $\Theta_{c,y}$ nor the neutral fermions.  Under $\T$ we have
\begin{eqnarray}
  \T: \Theta_{c,y} \rightarrow -\Theta_{c,y+1},  ~~~~
  \tilde \psi_y \rightarrow (-1)^y \tilde \psi_{y+1}.
\end{eqnarray}
Note that the neutral $\tilde \psi_y$ fields transform exactly as the original chiral domain wall electrons; see Eq.~\eqref{Tpsi}.  The coarse-grained dressed neutral fermions accordingly transform as
\begin{equation}
  \T: \tilde \Psi \rightarrow i \sigma^y \tilde \Psi.
\end{equation}
Finally, U(1)$_n$ sends
\begin{equation}
  {\rm U(1)}_n: \tilde \psi_y \rightarrow e^{i \alpha}\tilde \psi_y,  ~~~~\tilde \Psi \rightarrow e^{i \alpha}\tilde \Psi
\end{equation}
for arbitrary constant $\alpha$.  We can rephrase this last symmetry in more physical terms as follows.  In the extrinsic construction the Hamiltonian preserves U(1)$_n$ if electrons from the `$+$' and `$-$' quantum Hall edges are necessarily added or removed in tandem; this is clearly the case in both Eqs.~\eqref{Hc} and \eqref{NeutralFermionHopping}.  In the intrinsic construction U(1)$_n$ arises if $\psi_{+,y}$ and $\psi_{-,y'}$ electrons similarly appear pairwise as in Eqs.~\eqref{ChargeGapIntrinsic} and \eqref{NeutralFermionHoppingIntrinsic}.  Contrary to U(1) and $\T$, we can of course violate U(1)$_n$ without breaking any physical microscopic symmetries for our setup, though we assume for now that it persists and discuss various physical consequences.  

The CDL's charge gap immediately implies an \emph{incompressible} surface with activated diagonal electronic transport.  Note that the former property stands in stark contrast to the strictly 2D compressible composite Fermi liquid, despite their close relationship.  Both the spatially averaged electrical and thermal surface Hall conductances vanish due to $\T$ symmetry.  (Actually in our extrinsic construction $\sigma_{xy} = 0$ everywhere locally as well.)  Gapless neutral excitations supported by the CDL generate a $T^2$ contribution to the specific heat and `metallic' diagonal thermal transport that departs wildly from the Wiedemann-Franz law.  Together these characteristics readily distinguish the CDL from the usual gapless 3D TI surface.  Moreover, if descendant surface topological orders are eventually discovered experimentally such predictions could be relevant for intermediate temperature/voltage measurements.

\subsection{Preview of quasiparticles for composite Dirac liquid descendants}
\label{sec:cdlexcitations}

The CDL yields a number of general consequences for the proximate gapped phases---in particular the nature of deconfined excitations.  Following Ref.~\onlinecite{TeoKaneChains} our quasi-1D construction provides a convenient tool for identifying potential quasiparticles.  One can indeed anticipate the structure of excitations by examining individual domain walls.  For instance, solitons between degenerate minima of the $-\lambda \cos(8\Theta_{c,y})$ potential in the charge-sector Hamiltonian $H_c$ carry localized energy and thus comprise natural candidate quasiparticles.  When such solitons can also move \emph{between} domain walls without incurring an energy cost that grows with distance, they act as deconfined quasiparticles for the 2D surface.  
More generally, a \emph{fractional} deconfined quasiparticle always resides at the end of an `invisible' string operator that commutes with all measurements away from its endpoint.  Such strings must therefore incur no energy cost (i.e., commute with the Hamiltonian density), be locally neutral (i.e., commute with the charge density), and carry no other quantum numbers.  As an example, the appropriate string operators for all phases proximate to the CDL should not depend on $e^{iq\Phi_{c,y}}$ except near the endpoints, since this operator creates local charges.  

In fact, we have already encountered a simple example of such a string operator in Eq.~\eqref{eqn:dressedfermion}. The `dressed' neutral fermion $\tilde\Psi_y$ relates to the `bare' neutral fermion $\tilde \psi_y$ by the string $e^{2i \sum_{y' < y}(\Theta_{c,y'}+\Theta_{c,y'+1})}$.  While $\tilde \psi_y\sim e^{i\tilde\varphi_y}$ is locally defined in terms of bosonized fields, this operator by itself does \emph{not} create a physical excitation.  In the intrinsic construction the issue becomes immediately evident by comparing Eqs.~\eqref{IntrinsicElectrons} and \eqref{varphiIntrinsic}; $\tilde\psi_y$ clearly does not represent a local combination of electron operators.  A related problem occurs in the extrinsic constructions: there $\tilde \psi_y$ changes the total charge on individual quantum Hall fluids by fractional amounts, which can not happen.  In either case the $\Theta_{c,y}$ string heals these problems.  And since $\Theta_{c,y}$ is pinned in the CDL and all descendants that we examine, the string 
operator acquires an expectation value and becomes invisible in the sense defined above.  (As alluded to in Sec.~\ref{sec:universalcdl}, it is nevertheless still important to keep track of string operators to correctly obtain braiding statistics; see Sec.~\ref{Statistics} for more details.)  The dressed neutral fermion $\tilde\Psi_y$ readily hops both within and between domain walls via Eq.~\eqref{Hchains4} and consequently forms a deconfined surface excitation.  

Aside from neutral fermions, the charge-sector potential $-\lambda\cos(8\Theta_{c,y})$ suggests the presence of $e/4$ excitations as solitons where $\Theta_{c,y}\rightarrow \Theta_{c,y} + \pi/4$.  These solitons less obviously correspond to deconfined excitations, however, since their mobility between domain walls is unclear. Fortunately, one can build up operators that create well-separated $e/4$ dipoles along $y$ from the physical combinations
\begin{align}
  \mathcal{O}^{\rm ext}_{y+\frac{1}{2}}&\equiv e^{-i \phi_{1-,y+1}} e^{i \phi_{1+,y}}  \\
  &\sim e^{i (\Phi_{c,y}- \Phi_{c,y+1})/4}e^{i (-\Theta_{c,y}+\Theta_{c,y+1})/4}e^{i ( \tilde\varphi_y+ \tilde\varphi_{y+1})/4} \nonumber
\end{align}
in the extrinsic constructions and
\begin{align}
\mathcal{O}^{\rm int}_{y+\frac{1}{2}}&\equiv \psi_{-,y+1}^\dag\psi_y  \\
&\sim e^{i(\Phi_{c,y} - \Phi_{c,y+1})/4} 
e^{i(7\Theta_{c,y}+\Theta_{c,y+1})/4} e^{i(-3\tilde\varphi_y+\tilde\varphi_{y+1})/4} \nonumber
\end{align}
in the intrinsic case.  We emphasize that the bosonized form of $\mathcal{O}^{\rm ext}_{y+\frac{1}{2}}$ is general for a CDL obtained using \emph{any} auxiliary electronic $\nu=\pm 1/2$ quantum Hall fluids.  It only relies on the existence of $e/4$ excitations inside the quantum Hall domains, which is a generic feature of such states\cite{LevinStern}.  

From the dependence on the charge field $\Phi_{c,y}$, we see that either operator above creates an $e/4$ dipole across adjacent domain walls.  
The product 
\begin{equation}
  \Gamma_{e/4,y_2}^\dagger \Gamma_{e/4,y_1} \equiv \mathcal{O}_{y_2-1/2}\mathcal{O}_{y_2-3/2}\cdots \mathcal{O}_{y_1+3/2}\mathcal{O}_{y_1+1/2}
  \label{eover4dipole}
\end{equation} 
of such terms thus moves charge $e/4$ between arbitrary chains $y_1$ and $y_2$.  Importantly, the right-hand side yields a telescoping sum in the exponent such that the $\Phi_{c,y}$ dependence drops out away from the endpoints.  It is illuminating to decompose the extended $e/4$ dipole creation operator in Eq.~\eqref{eover4dipole} by writing
\begin{align}
  \Gamma_{e/4,y}  = \Psi_{e/4,y}\sigma_{y},
  \label{eover4}
\end{align}
where in the extrinsic construction
\begin{equation}
  \Psi_{e/4,y} = e^{i(\Phi_{c,y}  + \Theta_{c,y}-\tilde \varphi_{y})/4}
\end{equation}
removes the charge at the endpoint and
\begin{equation}
  \sigma_{y} = e^{-i \sum_{y' < y} \tilde \varphi_{y'}/2}
\end{equation}
is an electrically neutral string emanating downwards to $y = -\infty$. One can construct analogous (but slightly different) expressions for the intrinsic construction.  To avoid keeping track of the two sets of string operators, for simplicity we will hereafter focus on the extrinsic case when discussing excitations.  We have verified that consistent results for the quasiparticle content of the descendant states appear with either realization.

The question of whether deconfined $e/4$ excitations exist at the surface remains subtle.  Although the string $\sigma_y$ commutes with $H_c$ and is electrically neutral, it does not obtain an expectation value (unlike the $\Theta_{c,y}$ string for the neutral fermions) and moreover does not quite commute with $H_n$.  In the next section we nevertheless argue that a projected cousin of Eq.~\eqref{eover4} indeed describes deconfined $e/4$ excitations in the descendant T-Pfaffian state obtained by pairing the neutral fermions.  For the descendant Abelian topological order obtained by breaking $\T$ we will construct a modified string operator that \emph{does} have an expectation value indicating survival of deconfined $e/4$ quasiparticles there as well.  In the latter case we will also use such string operators to compute quasiparticle statistics.

\section{Composite Dirac liquid descendants: T-Pfaffian and 113 topological orders}
\label{descendants}

\subsection{Overview of  proximate phases}
 
Because the CDL is gapless, we can controllably access a number of interesting `nearby' phases by adding appropriate interactions.  We explore the following:

$(i)$ A symmetric gapped surface state that preserves the physical $\T$ and U(1) symmetries but breaks U(1)$_n$ conservation by Cooper pairing the neutral fermions.  This phase is accessed by perturbing $H_{\rm CDL}$ [Eq.~\eqref{HCDL}] with
\begin{eqnarray}
  H_\text{pair} &=& \sum_y (-1)^y\int_x \left[\tilde \Delta \tilde \psi_y \tilde \psi_{y+1}e^{-2i(\Theta_{c,y}-\Theta_{c,y+1})} + \mathrm{H.c.}\right]
  \nonumber \\
  &=& \sum_y (-1)^y \int_x \left(\tilde\Delta \tilde \Psi_y \tilde \Psi_{y+1} + \mathrm{H.c.}\right) 
  \nonumber \\
  &\sim& \int_{x,y}(\tilde\Delta \tilde \Psi_\uparrow \tilde \Psi_\downarrow + {\rm H.c.}).
  \label{HDelta}
\end{eqnarray}
Physically, the first line arises from processes such as $(\psi_{1+,y}^\dagger \psi_{1-,y+1})(\psi_{y+1}^\dagger\psi_y ) $ in the 331-based construction and $(\psi_{+,y}^\dagger \psi_{-,y+1})^2(\psi_y^\dagger \psi_{y+1})$ in the intrinsic construction.  On passing to the second line the strings cancel due to the pinning of $\Theta_{c,y}$.  Non-zero $\tilde \Delta$ drives the neutral fermions into a cousin of the familiar electronic TI superconducting surface state\cite{FuKane}---with the all-important difference that charge conservation persists.  We show in Sec.~\ref{TPfaffian} that the surface then realizes the symmetric T-Pfaffian state first captured by Bonderson et al.\cite{BondersonTO} and Chen et al.\cite{ChenTO}

$(ii)$ A $\T$-breaking gapped surface state obtained by adding a mass term for the neutral Dirac cone.  In coupled-chain language the mass arises upon dimerizing the neutral fermions via
\begin{eqnarray}
  &&H_\text{mag} = \sum_{y}\int_x \tilde m \left[\tilde \psi_y^\dag \tilde \psi_{y+1}e^{2i(\Theta_{c,y}+\Theta_{c,y+1})} + \mathrm{H.c.}\right]
  \nonumber \\
  &&= \sum_{y}\int_x \tilde m \left[\tilde \Psi_y^\dag \tilde \Psi_{y+1} + \mathrm{H.c.}\right] \sim \int_{x,y} \tilde m \tilde \Psi^\dagger \sigma^x \tilde \Psi.
  \label{Hm}
\end{eqnarray}
Notice that the above perturbation exhibits the same form as the $\tilde t$ term in Eq.~\eqref{Hchains4} but without the alternating $(-1)^y$ factor.  
Here the neutral fermions enter an analogue of the magnetically gapped TI surface, though $\sigma_{xy}$ vanishes because the Dirac cone no longer couples to the external electromagnetic field.  In Sec.~\ref{113} we identify this surface state with an Abelian topological order characterized by the K-matrix and charge vector
\beq
K = \begin{pmatrix} 1 & 3 \\ 3 & 1 \end{pmatrix},~~~~~ \vec q = \begin{pmatrix} 1 \\ 1\end{pmatrix}.
\label{K113}
\eeq

$(iii)$ `Nested' CDLs that exhibit a neutral Dirac cone coupled to gapped bosons with finer structure.  These symmetric gapless surface states follow by stripping off the `pseudocharge' associated with the emergent U(1)$_n$ symmetry, parallel to how we obtained a CDL from the original charged Dirac electrons.  Section \ref{NestedCDL} explores nested CDLs and their descendants, which interestingly includes the symmetric Pfaffian-antisemion phase uncovered through vortex condensation by Wang et al.\cite{WangTO} and Metlitski et al.\cite{MetlitskiTO} (and via Walker Wang models by Chen et al.\cite{ChenTO}).

\subsection{T-Pfaffian phase}
\label{TPfaffian}

One can anticipate the fate of the CDL perturbed by neutral-fermion pairing [Eq.~\eqref{HDelta}] by examining the boundary with a trivial, ferromagnetically gapped portion of the electronic TI surface.  
Very general arguments correctly capture the gapless modes appearing at the interface: The surface Hall conductance vanishes in the gapped CDL descendant whereas $\sigma_{xy} = \pm e^2/(2h)$ in the ferromagnetic region.  Hence the interface must support a chiral $\nu = 1/2$ charge mode.  Additionally, the paired neutral Dirac cone realizes a $\T$-invariant analog of a spinless 2D $p+ip$ superfluid and thereby contributes a gapless chiral Majorana mode.  We need only determine the relative chirality of these modes.

Note that the boundary of the Moore-Read state in a 2D system exhibits copropagating charge and Majorana modes; the closely related T-Pfaffian phase supports the same edge structure but with the opposite chirality for the Majorana sector.  At this point these phases comprise the natural candidate topological orders for our surface.  We can discriminate between these candidates as follows.  As discussed in Sec.~\ref{sec:universalcdl} the paired CDL has a vanishing surface \emph{thermal} Hall conductance $\kappa_{xy}$, while the ferromagnetic region has\cite{Yokoyama} $\kappa_{xy}=\frac{\pi^2 k_B^2}{3e^2}T\sigma_{xy}$---half the value of an ordinary $\nu = 1$ integer quantum Hall system.  To accommodate the jump in heat transport the interface must exhibit chiral central charge $c=\pm 1/2$, implying that the $\nu = 1/2$ charge mode and Majorana mode necessarily counterpropagate.\footnote{This conclusion also holds in the general case where the ferromagnetic regions carry $\sigma_{xy} = \pm (2n+1) e^2/(2h)$ for 
integer $n$.  The jump in electrical Hall conductance requires that the interface support $n$ $\nu = 1$ modes and one $\nu = 1/2$ mode, all of which copropagate.  The jump in thermal Hall conductance further implies a net chiral central charge of $c = n + 1/2$, which can only arise if the Majorana mode is counterpropagating. }  This logic identifies the symmetric gapped surface with the T-Pfaffian state\cite{BondersonTO,ChenTO}.  

\begin{figure}
\centering
\includegraphics[width=\columnwidth]{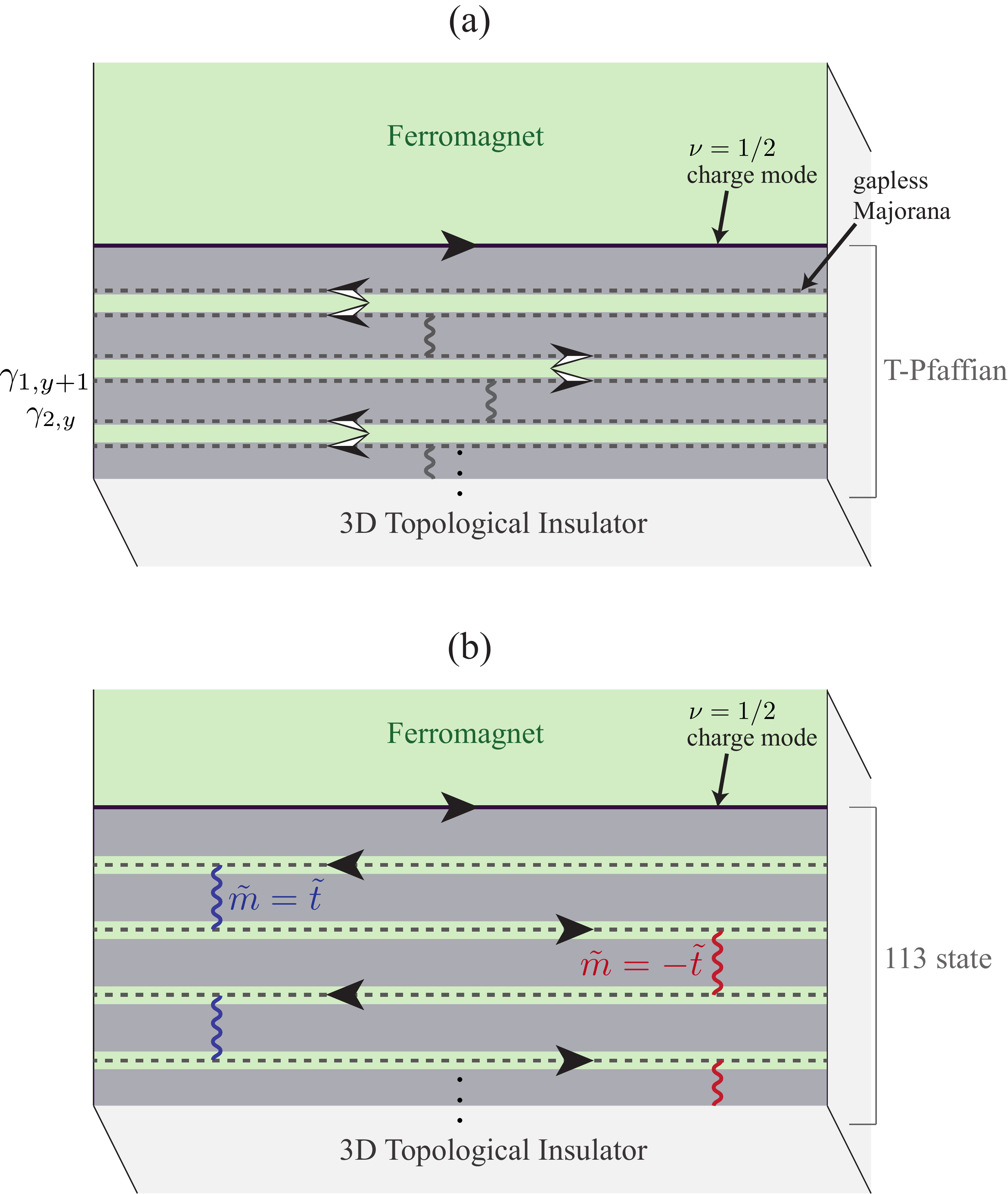}
\caption{(a) Interface between the symmetric T-Pfaffian surface topological order and a trivial magnetically gapped region in the extrinsic construction.  To construct the T-Pfaffian the neutral fermions from Fig.~\ref{ChargeGapFig}(b) are decomposed into a pair of Majorana modes $\gamma_{1,y}$ and $\gamma_{2,y}$.  The Majoranas then gap out by hybridizing with counterpropagating modes as indicated by wavy lines.  An `unpaired' chiral Majorana mode necessarily remains at the interface, together with a gapless $\nu = 1/2$ charge mode arising from the uppermost quantum Hall strip.  (b) Interface between the $\T$-breaking 113 surface topological order and a trivial magnetically gapped region.  Dashed lines represent neutral fermions dimerized in two possible ways depending on whether $\tilde m = \tilde t$ or $\tilde m = -\tilde t$.  In one case the interface supports only a $\nu = 1/2$ charge mode; in the other a counterpropagating $\nu = 1$ neutral mode coexists.  }
\label{EdgeFig}
\end{figure}

Our coupled-chain construction corroborates the above arguments.  The physics becomes particularly transparent upon specializing to the limit $\tilde \Delta = \tilde t$ and decomposing the dressed neutral fermions in terms of Majorana fields $\gamma_{1,y}$ and $\gamma_{2,y}$,
\begin{equation}
  \tilde \Psi_y = \gamma_{1,y}+i\gamma_{2,y}.
\end{equation}
Here the T-Pfaffian Hamiltonian $H_{\rm T\text{-}Pf} = H_{\rm CDL} + H_\text{pairing}$ simplifies to
\begin{eqnarray}
  H_{\rm T\text{-}Pf} &=& H_c + \sum_y (-1)^y \int_x \bigg{[} +i \hslash \tilde v_x \sum_{a = 1,2}\gamma_{a,y}\partial_x\gamma_{a,y}
  \nonumber \\
  &+& 4i\tilde\Delta\gamma_{2,y}\gamma_{1,y+1}\bigg{]}\label{eqn:htpfaffian}.
\end{eqnarray}
The last line implies the illuminating picture sketched in Fig.~\ref{EdgeFig}(a): the T-Pfaffian arises upon splitting each chiral neutral fermion into a pair of Majoranas, and then `democratically' annihilating them pairwise with Majoranas from neighboring domain walls.  This process preserves both $\T$ and U(1) yet eliminates all gapless excitations---at least away from interfaces where ungapped modes persist.  Figure~\ref{EdgeFig}(a) shows that the boundary with a trivial ferromagnetic region of the surface necessarily supports a $\nu = 1/2$ charge mode together with an `unpaired' counterpropagating Majorana, as deduced above.  One can understand the interface with a superconducting surface by nucleating a superconductor/ferromagnet/T-Pfaffian sequence and then shrinking the intervening ferromagnet; the 
Majorana from the former interface combines with that of the latter, yielding 
counterpropagating $\nu = 1/2$ and $\nu = 1$ modes as described in Ref.~\onlinecite{BondersonTO}.  

Section~\ref{sec:cdlexcitations} discussed the kinds of excitations expected in the CDL and descendant phases.  From that discussion, together with intuition from the more conventional superconducting TI surface, the T-Pfaffian's quasiparticle content becomes quite natural.  First, the surface clearly supports deconfined neutral fermions corresponding to Bogoliubov-de~Gennes quasiparticles in the paired-$\tilde\Psi_y$ condensate.  Given the nature of the charge gap it is also reasonable to expect an $e/4$ excitation associated with $\pi/4$ kinks in $\Theta_{c,y}$.  Recall from Sec.~\ref{sec:universalcdl} that a $\pi/4$ kink acts as an $h/(2e)$ superconducting vortex that yields a minus sign when encircled by a dressed $\tilde \Psi_y$ neutral fermion.  Thus in precise analogy with the superconducting TI surface,\cite{FuKane} localized $e/4$ charge binds a Majorana zero mode resulting in non-Abelian statistics.  See Appendix \ref{MajoranaAppendix} for an explicit solution of the zero mode in our quasi-1D 
formalism.  The neutral fermion and non-Abelian $e/4$ excitation generate all nontrivial quasiparticles present in the T-Pfaffian phase.  

In Sec.~\ref{sec:cdlexcitations} we constructed a candidate $e/4$ quasiparticle operator [Eq.~\eqref{eover4}] that was suggestively expressed as
\begin{align}
  \Gamma_{e/4,y} = e^{i(\Phi_{c,y} + \Theta_{c,y}-\tilde \varphi_y)/4} \sigma_y,~~~~ \sigma_y = e^{-i \sum_{y' < y} \tilde \varphi_{y'}/2}.\nonumber
\end{align}
This form is reminiscent of the $e/4$ quasiparticle operator for the edge of the Moore-Read state.  Indeed, very similar to the vortex field in the Ising theory describing the Moore-Read edge\cite{FendleyFisherNayak}, the non-local `twist operator' $\sigma_y$ effectively toggles boundary conditions on the neutral fermions:
\begin{equation}
  \sigma_{y_0}^\dagger(x_0) \tilde \Psi_{y<y_0}(x)\sigma_{y_0}(x_0) = -i(-1)^y \sgn(x-x_0)\tilde\Psi_{y<y_0}(x).
\end{equation}
(The twist operator does not affect neutral fermions with $y\geq y_0$.)  Despite this tantalizing correspondence, we know that $\Gamma_{e/4,y}$ does not represent the exact $e/4$ string operator.  The issue is that the non-local $\sigma_y$ piece commutes with $H_c$ and the neutral fermion pairing/interchain hopping terms (modulo a gauge transformation), but does \emph{not} commute with the neutral fermion kinetic energy along $x$.  Thus creating  well-separated $e/4$ charges via $\Gamma_{e/4,y}$ as written above unphysically raises the energy density in between the charge insertions.  Since the surface is gapped we can, however, project away these unphysical excitations by evolving the resulting state with $e^{-H_{\text{T-Pf}} \tau}$ for a finite time $\tau$ that scales as the inverse bulk gap.  Equivalently, replacing $\Gamma_{e/4,y}\rightarrow e^{H_{\text{T-Pf}}\tau} \Gamma_{e/4,y}e^{-H_{\text{T-Pf}}\tau}$ is expected to yield the desired operator corresponding to \emph{deconfined} $e/4$ quasiparticles.  
While we do not know how to explicitly evaluate this expression, we content ourselves with the physical picture provided here.  It would nevertheless be interesting to pursue a more exact treatment to put this discussion on more rigorous footing.

\subsection{$\mathcal{T}$-breaking 113 Abelian topological order}
\label{113}

Suppose now that we gap out the neutral fermions not by Cooper pairing, but instead by breaking our modified time-reversal symmetry $\T$ while preserving both U(1) and U(1)$_n$.  
Once we lift $\T$ a mass term for the CDL's neutral Dirac cone $H_\text{mag}  =\int_{x,y}\tilde m \tilde \Psi^\dagger \sigma^x \tilde \Psi$ becomes allowed and drives the surface into an Abelian topologically ordered state.  We can again anticipate the properties of this magnetically gapped CDL on general grounds.  As already noted the surface Hall conductance $\sigma_{xy}$ vanishes because of the Dirac cone's electrical neutrality.  In analogy with the ordinary ferromagnetically gapped TI surface, we can however assign the system a quantized neutral-fermion Hall conductance $\tilde \sigma_{xy}= \pm \tilde e^2/(2h)$, where $\tilde e$ is the emergent `pseudocharge' associated with U(1)$_n$ symmetry.  An immediate physical implication is that a domain wall at which $\tilde m$ changes sign binds a gapless \emph{neutral} chiral fermion mode.  Also in analogy with the usual ferromagnetic TI surface, the gapped neutral Dirac cone yields a thermal Hall conductance with the same magnitude as for the massive charged 
Dirac cone, $\kappa_{xy}=\frac{\pi^2 k_B^2}{3\tilde e^2}T\tilde\sigma_{xy}$.  

These properties allow us to deduce the structure of the magnetically gapped CDL/ferromagnet interface.  First the jump in electrical Hall conductance dictates that the border must bind a chiral $\nu = 1/2$ charge mode.  Depending on the relative sign of the Dirac cone masses in each region, the thermal Hall conductance either $(i)$ jumps by a discrete amount that requires a chiral central charge $c = 1$ for the interface or $(ii)$ matches on both sides---in which case the chiral central charge must vanish.  Moreover, we can pass between these two cases by simply changing the sign of the neutral fermion mass $\tilde m$.  It follows that in $(i)$ the minimal edge structure consists of a lone $\nu = 1/2$ charge mode whereas case $(ii)$ features an additional counterpropagating gapless neutral fermion.  The simplest topological order consistent with these boundary modes corresponds to the 113 state with K-matrix given in Eq.~\eqref{K113}.  We confirm this identification below through a detailed analysis of the 
bulk quasiparticles supported by the magnetically gapped CDL.  

The gapless modes at the interface follow explicitly from our quasi-1D approach in the special case where $\tilde m = \tilde t$.  Here the Hamiltonian reduces to
\begin{eqnarray}
  H_{113} &=& H_c + \sum_y  \int_x [ +i(-1)^y \hslash \tilde v_x \tilde \Psi_y^\dagger \d_x \tilde \Psi_y 
  \nonumber \\
  &+& 2\tilde m (\tilde\Psi_{2y-1}^\dagger \tilde \Psi_{2y} + {\rm H.c.})].
\label{eqn:h113}
\end{eqnarray}
From the last line we see that the neutral fermions gap out by dimerizing such that chain 1 hybridizes with 2; 3 hybridizes with 4; etc.  Of course the shifted dimerization is also possible and arises upon taking $\tilde m = -\tilde t$.  Figure \ref{EdgeFig}(b) illustrates the boundary with a ferromagnet in both dimerizations, which indeed reproduce the two possible edge structures identified using general arguments above.  

Other interfaces are also interesting and can be understood similarly.  For instance, a superconductor/113 boundary harbors a chiral Majorana mode in addition to a $\nu = 1/2$ mode.  Furthermore, the interface separating the T-Pfaffian and 113 surface topological orders also hosts a chiral Majorana mode (with no other structure)---precisely as in the boundary between the usual superconducting and ferromagnetic TI surface.  The fact that these interfaces look identical reflects the close analogy between the gapped phases accessible from the CDL and the original charged Dirac cone.

\subsubsection{Explicit calculation of 113 topological order}
\label{Statistics}

To verify our claim that the magnetically gapped CDL realizes the topological order of an Abelian 113 state we will now explicitly calculate the charge and statistics of deconfined quasiparticles living at the surface.
We begin by recalling the elementary operator introduced in Sec.~\ref{sec:cdlexcitations} that allows propagation of $e/4$ excitations between domain walls,
\begin{equation}
  \mathcal{O}_{y+\frac{1}{2}}\sim e^{i (\Phi_{c,y}- \Phi_{c,y+1})/4}e^{i (-\Theta_{c,y}+\Theta_{c,y+1})/4}e^{i ( \tilde\varphi_y+ \tilde\varphi_{y+1})/4}.
  \nonumber
\end{equation}
This operator not only moves charge $e/4$ from $y$ to $y+1$, but also adds pseudocharge $\tilde e/4$ to \emph{both} domain walls. 
Accordingly, a product $\cdots \mathcal{O}_{y-1/2}\mathcal{O}_{y+1/2}\mathcal{O}_{y+3/2}\cdots$ that creates a well-separated $e/4$ dipole leaves behind a trail of $\tilde e/2$ psuedocharges along the string connecting the endpoints.  Preservation of U(1)$_n$ symmetry implies conservation of pseudocharge; the string thus carries nontrivial quantum numbers and is clearly not invisible.  We can, however, modify the string in a simple way to rectify this problem.  

Observe that there exists a second operator that also transfers charge $e/4$ between domain walls:
\begin{align}
\mathcal{O}'_{y+1/2}&= \mathcal{O}_{y+1/2}^{-3}\psi^\dagger_{y+1}\psi_y.
\end{align}
We emphasize that $\psi_y$ represents a physical magnetic domain-wall electron operator, as opposed to a neutral fermion $\tilde \psi_y$.  The operator $\mathcal{O}'_{y+1/2}$ \emph{removes} pseudocharge $3\tilde e/4$ from both domain walls $y$ and $y+1$.  By suitably assembling strings from products of $\mathcal{O}$ and $\mathcal{O}'$ one can thus move charge $e/4$ across arbitrary domain walls without leaving behind a net pseudocharge trail.  
One such product reads
\begin{align}
  \hat v_1= \cdots\mathcal{O}_{y-1/2} \mathcal{O}_{y+1/2} \mathcal{O}'_{y+3/2}\mathcal{O}_{y+5/2} \mathcal{O}_{y+7/2} \mathcal{O}'_{y+9/2}\cdots \label{eqn:113string},
\end{align}
while another, $\hat v_2$, arises by replacing $\mathcal{O}' \leftrightarrow \mathcal{O}$.  Apart from carrying no net pseudocharge the strings involved obtain an expectation value and hence are invisible.  (The nonzero expectation value can be seen by expressing the neutral fermion Hamiltonian in bosonized language.)  
This demonstrates that the magnetically gapped CDL supports deconfined $e/4$ excitations, which come in two flavors due to the inequivalent strings $\hat v_1$ and $\hat v_2$.  Physically, starting from one $e/4$ flavor we obtain the other by fusing with a neutral fermion $\tilde \Psi_y$.  

We can further leverage these string operators to deduce the exchange statistics of the elementary $e/4$ excitations.  To do so we extract operators $\hat h_{1,2}$ that create extended `horizontal' $e/4$ dipoles within a given domain wall; braiding properties then follow upon commuting with the `vertical' dipole operators $\hat v_{1,2}$.  
As a first step we terminate the $\hat v_{1,2}$ strings at some domain wall $y_0$. The endpoints of the terminated string create both charge and pseudocharge (unlike the intervening string) and thus neither acquires an expectation value nor commutes with other string operators.  Given Eq.~\eqref{eqn:113string} the precise form of the operator that terminates a string will clearly depend on the value of $y_0$.  However, one can verify that the statistical phases obtained by braiding quasiparticles does not depend on this choice. 

As a specific example, consider vertical dipole operators $\hat v_{1,2}$ defined with a string emanating from $y = -\infty$ and terminating at a domain wall $y_0$ such that
\small
\begin{align}
 \hat v_{1} \rightarrow \cdots \mathcal{O}_{y_0-7/2} \mathcal{O}_{y_0-5/2} \mathcal{O}'_{y_0-3/2}\mathcal{O}_{y_0-1/2}\\
 \hat v_{2} \rightarrow \cdots \mathcal{O}'_{y_0-7/2}\mathcal{O}'_{y_0-5/2} \mathcal{O}_{y_0-3/2}\mathcal{O}'_{y_0-1/2}.
\end{align}
\normalsize
For odd $y_{0}$ the endpoint operators are respectively
\begin{align}
&\xi_{1,y_{0}}^\dagger =  e^{ i (-\Phi_{c,y_{0}}+\Theta_{c,y_{0}} + \tilde \varphi_{y_{0}})/4}
\label{xi1} \\
&\xi_{2,y_{0}}^\dagger =  e^{ i (-\Phi_{c,y_{0}}-7\Theta_{c,y_{0}} -3 \tilde \varphi_{y_{0}})/4}.
\end{align}
Terminating at $y'_{0} = y_{0}+1$ instead yields\footnote{The second set of operators involves \emph{two} domain walls, which are both part of the same, enlarged unit cell in this dimerized phase.  It is necessary to include these extra pieces because the string operators only acquire expectation values upon crossing an entire unit cell due to the dimerization. However, as we demonstrate explicitly this short-scale structure does not affect the universal (long wavelength) properties such as spin.}  
\begin{align}
&\xi_{1,y'_{0}}^\dagger =  e^{ i (-\Phi_{c,y'_{0}}+\Theta_{c,y'_{0}} + \tilde \varphi_{y'_{0}} +2 \tilde \varphi_{y'_{0}-1})/4}
\\
&\xi_{2,y'_{0}}^\dagger =  e^{ i (-\Phi_{c,y'_{0}}-7\Theta_{c,y'_{0}}  -3 \tilde \varphi_{y'_{0}} -6 \tilde \varphi_{y'_{0}-1})/4}.
\label{xi4}
\end{align}
Despite the rather different form of these endpoint operators, both satisfy commutation relations
\begin{align}
  \xi_{i,y} ^\dagger(x) \xi_{i,y}^\dagger(x')= \xi_{i,y}^\dagger(x')\xi_{i,y}^\dagger (x) \exp \frac{i \pi \text{sgn}(x'-x)}{8}
  \label{eqn:113spins} ,
\end{align}
indicating that the $e/4$ quasiparticles carry spin $e^{i\pi/8}$.  Note that these results apply for a particular sign of the mass $\tilde m$ in Eq.~\eqref{eqn:h113}; the opposite sign arises upon applying $\T$ and thus yields spins $e^{-i\pi/8}$.  

Equations \eqref{xi1} through \eqref{xi4} allow us to construct operators $\hat h_{1,2}$ that create extended horizontal $e/4$ dipoles.  We will focus for simplicity on  dipoles created by
\begin{align}
  \hat h_1 &=\xi^\dagger_{1,y_{0}}(-\infty)\xi_{1,y_{0}}(\infty) \nonumber \\
&= e^{i\int_{-\infty}^{\infty} dx \partial_x (-\Phi_{c,y_{0}} + \Theta_{c,y_{0}}+\tilde\varphi_{y_{0}})/4}
\end{align}
and likewise for $\hat h_2$.  Note that the right side involves only densities and hence represents a manifestly physical operator.
Now consider the operator
\begin{equation}
  \hat v_a^\dagger \hat h_b^\dagger \hat v_a \hat h_b \equiv e^{i \theta_{ab}}.
  \label{fullbraid}
\end{equation}
The sequence on the left-hand side simulates a full braid of an $e/4$ of flavor $a$ around an $e/4$ of flavor $b$, resulting in a statistical phase $\theta_{ab}$. 
Since the only part of the vertical strings that non-trivially commute with the $\xi_{i,y_0}$ operators are
\begin{align}
&\hat{v}_1 = \cdots e^{ i \tilde \varphi_{y_0}/2} \cdots\\
&\hat{v}_2 = \cdots e^{ -i 3 \tilde \varphi_{y_0}/2}\cdots,
\end{align}
the statistical angles are straightforward to evaluate.  We find 
\begin{eqnarray}
  \theta_{11} = \theta_{22} = \frac{-\pi}{4},~~~~~~\theta_{12} = \theta_{21} = \frac{3\pi}{4}.
  \label{BraidAngles}
\end{eqnarray}

Given the statistics for a full braid involving just one $e/4$ flavor (say, $\theta_{11}$) we can actually intuitively extract the other statistical phases above.  We simply need to recall that attaching a neutral fermion produces the other $e/4$ flavor, and that a neutral fermion acquires a minus sign when encircling an $e/4$ charge.  It follows that the same phase arises upon braiding either $e/4$ around another $e/4$ of the same type (i.e., $\theta_{11} = \theta_{22}$), while their mutual statistics contains an extra factor of $\pi$ (i.e., $\theta_{i\neq j} = \theta_{ii} + \pi$)---both in harmony with Eq.~\eqref{BraidAngles}.
The spins computed earlier as well as these statistical phase factors precisely match those for $e/4$ excitations in a topological phase with K-matrix and charge vector given in Eq.~\eqref{K113}, thereby justifying our identification of the magnetically gapped CDL with the 113 state.

\section{Nested composite Dirac liquids}
\label{NestedCDL}

We now shift gears and study additional states that arise if the neutral fermions forming the CDL do not trivially gap out, but rather strongly interact yielding `nested' CDLs and descendants thereof.  To this end we invoke an algorithm analogous to that used for stripping the electric charge from the original surface Dirac cone.  Let us briefly recapitulate the algorithm.  Starting from an array of charged electron modes $\psi_y$ [Fig.~\ref{Chains}(a)] we deposited into each magnetic domain auxiliary $\nu = \pm 1/2$ quantum Hall strips [Fig.~\ref{ChargeGapFig}(a)] that facilitated opening a charge gap.  We then obtained an array of electrically neutral fermions $\tilde \psi_y$ [Fig.~\ref{ChargeGapFig}(b)] with opposite chirality relative to the $\psi_y$ modes.  

It again proves very helpful to view $\tilde \psi_y$ as carrying `pseudocharge' associated with an emergent U(1)$_n$ symmetry---which we wish to now strip off in a similar vein.  Imagine, then, depositing quantum Hall strips that allow us to generate a \emph{pseudocharge} gap.  (We require that the quantum Hall strips support neutral `edge semions' that play the role of the $\nu = 1/2$ charge modes in our previous CDL constructions.)  When the dust settles gapless fermions $\tilde \psi_y'$ emerge with reversed chirality relative to $\tilde \psi_y$.  The $\tilde \psi'_y$ fermions carry neither electrical charge nor pseudocharge; $\T$-preserving tunneling amongst these modes yields a second-generation neutral Dirac cone.  This gapless state comprises a nested CDL that serves as a mother theory for a new set of non-trivial gapped surface phases.

\subsection{Pfaffian-antisemion phase}
\label{PfaffianAntisemion}

The most interesting descendant phase arises upon Cooper pairing the second-generation neutral Dirac cone, resulting in a $\T$-invariant, electric-charge conserving gapped surface for the electronic TI.  We can fruitfully view this state as a T-Pfaffian formed by neutral $\tilde \psi_y$ fermions---rather than the original electrons $\psi_y$---that coexists with a residual $\nu = 1/2$ charge sector inherited from the original CDL.  Many properties immediately follow from this vantage point.  Notably, the boundary with a trivial ferromagnetic surface region hosts oppositely oriented \emph{neutral} $\nu = 1/2$ and Majorana modes (from the $\tilde \psi_y$ T-Pfaffian) together with a $\nu = 1/2$ charge mode (from the residual charge sector).  The two $\nu = 1/2$ modes necessarily counterpropagate; this follows from the reversed chirality of the $\psi_y$ and $\tilde \psi_y$ fermions or, alternatively, from the difference in thermal Hall conductance across the interface.  In other words the interface looks identical to the Moore-
Read edge combined with a backward propagating $\nu = 1/2$ neutral mode as Fig.~\ref{SummaryFig} illustrates.  This result identifies the pairing-gapped nested CDL with the Pfaffian-antisemion symmetric surface phase obtained by completely different means in Refs.~\onlinecite{WangTO,ChenTO,MetlitskiTO}.  Quite remarkably, our formalism thus reveals a hierarchical relation between the T-Pfaffian and Pfaffian-antisemion topological orders discussed previously and captures both on equal footing.

\subsection{$\T$-breaking 331-antisemion Abelian topological order}
\label{331antisemion}

We can also gap the nested CDL by adding a magnetic mass term for the second-generation neutral Dirac cone, giving way to a $\T$-violating Abelian surface topological order.  The properties can again be immediately deduced by analogy with the descendants of the original CDL.  In particular, one can view the gapped phase as a 113 state formed by $\tilde \psi_y$ fermions together with a residual $\nu = 1/2$ charge sector.  Thus the boundary with a trivial ferromagnet necessarily harbors a neutral $\nu = 1/2$ mode from the former and a backwards-propagating $\nu = 1/2$ charge mode from the latter; a backwards-propagating $\nu = 1$ mode may also arise depending on the sign of the magnetic mass.  This edge structure is remarkably similar to that of a 331 bilayer state together with a counterpropagating $\nu = 1/2$ neutral mode, and suggests that the K-matrix and charge vector for the surface topological order are given by
\beq
K = \begin{pmatrix} 3 & 1 & 0 \\ 1 & 3 & 0 \\ 0 & 0 & -2 \end{pmatrix},~~~~~ \vec q = \begin{pmatrix} 1 \\ 1 \\ 0 \end{pmatrix}.
\label{K331}
\eeq
Section \ref{2Dconnection} provides strong support for this claim.
Thus we label the magnetically gapped nested CDL as the `331-antisemion' phase.  Interestingly, the 331 state can be understood as a relative of the Moore-Read state in which spinful composite fermions undergo triplet pairing.\cite{ReadGreen}  Thus the appearance of 331 physics at the surface is rather natural given the proximity to the Pfaffian-antisemion phase.  We further explore such correspondences in the next section.

\subsection{Higher nestings}

One can of course continue the procedure of nesting the CDL ad infinitum---though higher nestings can always be reduced to either the first- or second-generation CDLs examined so far.  It is again instructive to consider the boundary between a gapped descendant of a higher-nested CDL and a trivial ferromagnet.  The key principle is that subsequent nestings simply reverse the orientation of the gapless modes from the previous generation and contribute an extra neutral $\nu = 1/2$ mode.  Thus two repeated nestings add a set of neutral, \emph{counterpropagating} $\nu = 1/2$ modes to the previous interface structure.  Condensing a semion-antisemion pair can gap out these additional interface modes and confine the extra anyon species at the surface arising from two repeated nestings.  In this sense our construction identifies only two `fundamental' CDLs whose gapped descendants do not connect through simple 
condensation transitions.

\renewcommand{\i}{\mathrm{i}}

\section{Connection to 2D phases}
\label{2Dconnection}

Upon breaking $\T$ symmetry \emph{all} of the descendant topological orders captured so far---including the T-Pfaffian and Pfaffian-antisemion---can arise in strictly 2D systems and hence be `peeled off' of the surface.  Here we show precisely how this transpires beginning from the parent CDLs.  Along the way we will gain new insight into precisely why the surface physics with $\T$ is special and relate the CDL descendants to paired phases of 2D composite Fermi liquids.  This exercise also reveals an alternative (and somewhat surprising) way of viewing the Abelian surface topological orders that establishes a more intuitive relationship to the proximate non-Abelian states.  

As a primer we briefly review the phases accessible from the standard 2D composite Fermi liquid arising in spin polarized, filling factor $\nu = 1/2$ systems.\cite{ReadGreen,GreiterPaired,Milovanovic,ReadRezayi}  For this discussion it is convenient to decompose the electrons $c_{\bf r}$ in terms of partons via 
\begin{equation}
  c_{\bf r} = f_{\bf r} b_{\bf r},
\end{equation}
where $b_{\bf r}$ represents a boson that carries all the electric charge while $f_{\bf r}$ denotes a neutral composite fermion that sees zero magnetic field on average.  (In the physical Hilbert space the bosons and fermions bind to one another.)  The composite Fermi liquid arises if the former realizes a $\nu = 1/2$ bosonic quantum Hall state while the latter forms a gapless Fermi sea.  Putting the composite fermions into a `weakly paired' $p\pm ip$ state gaps the Fermi surface, resulting in an incompressible phase in which the edge hosts a $\nu = 1/2$ charge mode (from the $b$'s) and a chiral Majorana mode (from the $f$'s).  Their relative orientation depends on whether $p+ip$ or $p-ip$ pairing occurs; the Moore-Read phase arises if these modes copropagate while otherwise the T-Pfaffian emerges.  

Starting from either non-Abelian phase, we can obtain a nearby Abelian topological order at the same filling factor by reducing the composite-fermion chemical potential until it drops below the band bottom.  In the process the weakly paired composite Fermi sea undergoes a phase transition into a `strong pairing' state arising from tightly bound composite fermion pairs.\cite{HalperinQH}  The chiral Majorana mode from the non-Abelian phases then disappears, yielding an Abelian state with a lone $\nu = 1/2$ charge mode at the boundary.  Because the fermionic and bosonic partons are tied together in the physical Hilbert space, we can view this phase as a \emph{bosonic} $\nu = 1/2$ quantum Hall state built from charge-$2e$ electron pairs.  The K-matrix and charge vector for the Abelian topological order read
\begin{equation}
  K = 8,~~~~~~q = 2.
  \label{StrongPairing}
\end{equation} 
Excitations thus carry charge $2 l/8$ and spin $e^{i \pi l^2/8}$ for integer $l$.  In particular, there exists an $e/4$ quasiparticle with spin $e^{i \pi/8}$; a fractionalized charge-$e$ \emph{boson} that yields a minus sign when braided all the way around this $e/4$; and a local charge-$2e$ boson reflecting the elementary electron pairs.  Fermionic quasiparticles are simply absent in the K-matrix description, though electrons still of course exist at finite energies; for a recent example of this phenomenon in the literature, see Ref.~\onlinecite{Cano}.

\begin{figure}
\centering
\includegraphics[width=\columnwidth]{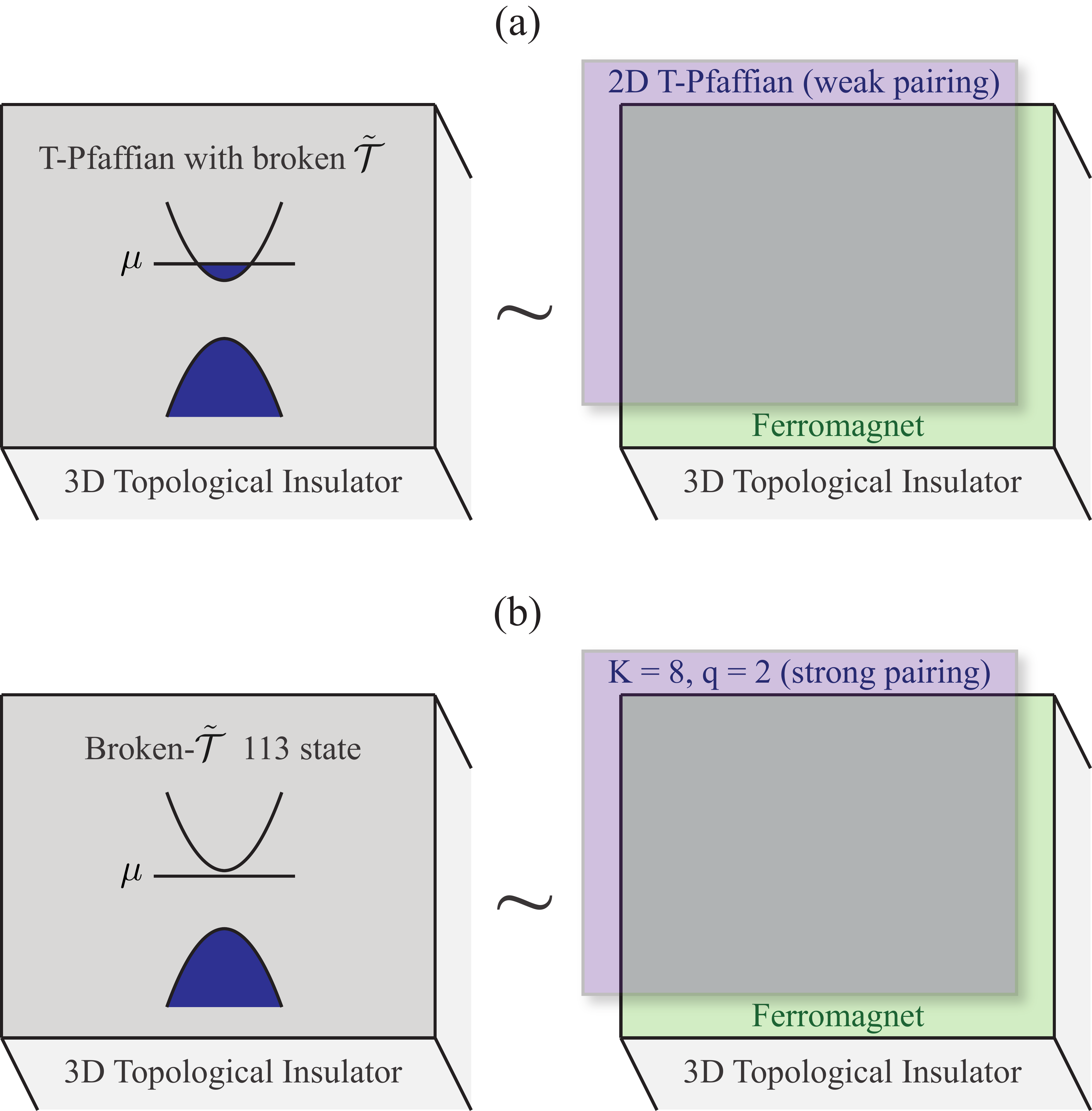}
\caption{Connection between CDL descendants in the absence of $\T$ symmetry and strictly 2D topological orders.  (a) With broken $\T$ the pairing-gapped CDL reduces to a trivial ferromagnetically gapped surface together with a 2D T-Pfaffian arising from a weakly paired composite Fermi liquid.  (b) Conversely, the magnetically gapped CDL is equivalent to a ferromagnetic surface coexisting with a strongly paired composite Fermi liquid state with K-matrix $K = 8$ and charge vector $q = 2$. }
\label{2Dfig}
\end{figure}

Let us now endeavor to connect these 2D composite Fermi liquid descendants to the electronic TI surface with broken $\T$ symmetry.  As a first step we start from the original (un-nested) CDL and add a `large' mass $\tilde m$ for the neutral Dirac cone.  Assuming the neutral fermion chemical potential intersects the upper half of the massive Dirac cone as shown in Fig.~\ref{2Dfig}(a), we can safely project away the lower half.  The remaining gapless neutral fermions then closely emulate the 2D composite Fermi liquid's Fermi sea.  Using adiabatic continuity of our results from Sec.~\ref{TPfaffian} we know that incorporating pairing $\tilde \Delta$ as in Eq.~\eqref{HDelta} drives the surface into a T-Pfaffian with broken $\T$---precisely as for a weakly paired 2D composite Fermi liquid with appropriate chirality.  It follows that, on breaking $\T$, the electronic TI surface in the T-Pfaffian phase realizes the same universal physics as a trivial magnetically gapped surface superimposed with a T-Pfaffian from a 
strictly 2D system.  Figure \ref{2Dfig}(a) illustrates the correspondence.  (The magnetically gapped surface merely cancels the Hall conductance from the 2D system so that the response properties match.)  Conversely, with $\T$ symmetry intact the projection above becomes invalid; one can then not strip the T-Pfaffian from the surface for exactly the same reason that the time-reversal-invariant superconducting TI surface can not separate from the bulk.  

Suppose next that, as sketched in Fig.~\ref{2Dfig}(b), the chemical potential descends into the mass gap for the CDL's neutral Dirac cone.  Given the correspondence just identified, it is natural to assume that the Abelian phase for a strongly paired 2D composite Fermi liquid takes over.  However, we face a conundrum here: Again by adiabatic continuity our results from Sec.~\ref{113} indicate that the surface realizes an (electronic) 113-state topological order with
\beq
K_{113} = \begin{pmatrix} 1 & 3 \\ 3 & 1 \end{pmatrix},~~~~~ \vec q_{113} = \begin{pmatrix} 1 \\ 1 \end{pmatrix},
\label{K113b}
\eeq
whereas the strong pairing state corresponds to the (bosonic) K-matrix and charge vector given in Eq.~\eqref{StrongPairing}.  Moreover, a deconfined neutral fermion clearly exists for the electronic TI surface whereas the latter state admits \emph{no} fermionic quasiparticles as noted above.  How are these apparently contradictory facts reconciled?

A basis change proves extremely illuminating.  We use the known fact\cite{WenBook} that $K$, $\vec q$ and $K'$, $\vec q'$ describe the same topological order if there exists an integer-valued matrix $W$ with determinant $\pm 1$ that satisfies $W^T K W = K'$ and $W^T q = q'$.  Defining one such matrix
\begin{equation}
  W = \begin{pmatrix} 1 & 0 \\ -3 & 1 \end{pmatrix},
\end{equation}
we obtain
\begin{equation}
  W^T K_{113}W = \begin{pmatrix} -8 & 0 \\ 0 & 1 \end{pmatrix},~~~~~~~W^T \vec q_{113} = \begin{pmatrix} -2 \\ 1 \end{pmatrix}.
  \label{K113c}
\end{equation}
It follows that the 113 state arises from the strongly paired composite Fermi liquid upon adding an \emph{unfractionalized} $\nu = 1$ integer quantum Hall state!  Topological orders described by Eqs.~\eqref{StrongPairing} and \eqref{K113b} actually exhibit the same anyonic content.\footnote{Technically they describe time-reversed conjugates of each other as written because of the relative minus signs in Eqs.~\eqref{StrongPairing} and \eqref{K113c}.}  In particular, the neutral fermion for our surface arises upon binding a physical electron to the charge $-e$ boson evident from the strong pairing state.  One can thereby indeed distill the magnetically gapped CDL into a trivial ferromagnetic surface superimposed with a 2D strongly paired $\nu = 1/2$ quantum Hall state as summarized in Fig.~\ref{2Dfig}(b).  

Analogous results hold for the nested CDL descendants.  Consider a 2D system composed of electrons forming a composite Fermi liquid together with neutral excitons that realize a $\nu = -1/2$ quantum Hall state.  The 2D Pfaffian-antisemion phase emerges upon weakly pairing the composite fermions such that the edge supports copropagating $\nu = 1/2$ charge and Majorana modes along with a counterpropagating $\nu = 1/2$ neutral mode contributed by the excitons.  Following precisely the same logic as before, entering a strongly paired state by lowering the composite-fermion chemical potential yields an Abelian \emph{bosonic} topological order with
\beq
K = \begin{pmatrix} 8 & 0 \\ 0 & -2 \end{pmatrix},~~~~~ \vec q = \begin{pmatrix} 2 \\ 0 \end{pmatrix} .
\label{StrongPairing2}
\eeq
The lower entries simply encode the neutral exciton quantum Hall sector.  

Imagine adding a large $\T$-breaking mass now to the nested CDL.  Provided the chemical potential remains far from the lower half of the Dirac cone we can again capture the low-energy surface physics by projecting onto the upper branch.  Doing so effectively maps the problem onto that of a 2D composite Fermi sea.  With the Fermi level outside of the mass gap, pairing drives the nested Dirac cone into the Pfaffian-antisemion phase in agreement with this mapping (recall Sec.~\ref{PfaffianAntisemion}).  Curiously, the arguments from Sec.~\ref{331antisemion} imply that placing the chemical potential within the mass gap generates topological order with
\beq
  K_{331\text{-as}} = \begin{pmatrix} 3 & 1 & 0 \\ 1 & 3 & 0 \\ 0 & 0 & -2 \end{pmatrix},~~~~~ \vec q_{331\text{-as}} = \begin{pmatrix} 1 \\ 1 \\ 0\end{pmatrix},
\label{K331b}
\eeq
which appears at odds with Eq.~\eqref{StrongPairing2}.  A basis change again resolves the issue.  In this case the matrix
\begin{equation}
  W = \begin{pmatrix} -3 & 1 & 1 \\ 1 & -1 & 0 \\ 4 & -1 & -1 \end{pmatrix}
\end{equation}
allows us to write
\begin{equation}
  W^T K_{331\text{-as}}W = \begin{pmatrix} -8 & 0 & 0 \\ 0 & 2 & 0 \\ 0 & 0 & 1 \end{pmatrix},~~~~~ W^T \vec q_{331\text{-as}} = \begin{pmatrix} -2 \\ 0 \\ 1 \end{pmatrix}.
\end{equation}
Topological orders described by Eqs.~\eqref{StrongPairing2} and \eqref{K331b}, remarkably, also therefore differ by an unfractionalized integer quantum Hall state.  Note that since we did not explicitly derive Eq.~\eqref{K331b} (unlike for the 113 topological order) the correspondence obtained here is very reassuring and serves as strong evidence that we have correctly identified the physics.
To summarize, in the absence of $\T$ symmetry the gapped descendants of the nested CDL indeed reduce to a trivial ferromagnetic surface coexisting with weak or strong pairing 2D quantum Hall phases---just like the original CDL.

\section{Application to bosonic surface topological order}
\label{BosonicTO}

The quasi-1D decomposition developed in this paper allows one to controllably capture surface topological orders from more than just the familiar 3D electronic TI.  
To illustrate the generality of our approach, we now apply similar methods to \emph{bosonic} 3D TIs discovered by Vishwanath and Senthil\cite{AshvinSenthil}.  We will treat two distinct bosonic topological phases---one protected by time-reversal and charge conservation symmetries, the other just time reversal---and recover the surface topological orders obtained by different means in Refs.~\onlinecite{AshvinSenthil,BurnellChen,ScottLesik}.

\subsection{Bosonic topological insulator protected by time reversal and charge conservation}

We first consider a 3D bosonic TI that, in the language of Ref.~\onlinecite{AshvinSenthil}, is protected by U(1)$\rtimes Z_2^T$ symmetry corresponding to boson number conservation and time reversal.  A magnetically gapped surface for such a phase carries a vanishing thermal Hall conductance $\kappa_{xy} = 0$ but nontrivial charge  conductance $\sigma_{xy} = \pm e^2/h$.  As for the electronic TI the latter value is anomalous in that a purely 2D bosonic system with this charge Hall conductance necessarily supports fractionalized excitations---whereas the surface does not.  Magnetic domain walls must therefore bind unfractionalized gapless modes that are nonchiral (because $\kappa_{xy} = 0$) but accommodate a jump in $\sigma_{xy}$ of $\pm 2e^2/h$.  Gapless modes described by a K-matrix $K = \sigma^x$ and charge vector $\vec q = (1,1)^T$ satisfy both constraints.  

Suppose that, analogously to the electronic case, we abandon local time-reversal invariance in favor of $\T$ symmetry and pattern an array of magnetic domain walls on the surface similar to Fig.~\ref{Chains}(a).  We thus obtain a set of $K = \pm \sigma^x$ modes with alternating chirality.  Precisely this `network model' was analyzed in Ref.~\onlinecite{AshvinSenthil} when constructing a gapless surface field theory for the bosonic topological insulator.  Here we will instead judiciously decorate the surface to controllably gap out the domain-wall modes and enter a $\T$-invariant topologically ordered surface phase.  With this objective in mind, fractionalized 2D bosonic quantum Hall systems with $K = \pm 2\sigma^x$ and $\vec q = (1,1)^T$ constitute ideal auxiliary systems to add into the domains.  These systems carry $\kappa_{xy} = 0$ and $\sigma_{xy} = \pm e^2/h$; hence they can cancel the charge Hall conductance from the magnetic domains without altering the trivial surface thermal Hall conductance.  Gapless domain-wall modes need no longer appear in the decorated surface as we now demonstrate.

Contrary to the electronic TI surface, perturbations acting within individual domain walls suffice to completely open a gap (because both $\sigma_{xy}$ \emph{and} $\kappa_{xy}$ now vanish).  Including the upper and lower edge states from the adjacent bosonic quantum Hall fluids, the full set of gapless modes at a particular domain wall is described by
\begin{align}
 &K=\begin{pmatrix}
  0&-2&0&0&0&0\\
  -2&0&0&0&0&0\\
  0&0&0&1&0&0\\
  0&0&1&0&0&0\\
  0&0&0&0&0&-2\\
  0&0&0&0&-2&0\\
 \end{pmatrix}
 \label{eqn:sigmax}
 ,~~~~~~\vec q=\begin{pmatrix}1\\1\\1\\1\\1\\1\end{pmatrix}.
\end{align}
We denote the fields corresponding to this K-matrix by
\begin{align}
\vec \phi = (\phi_{1+}, \phi_{2+}, \varphi_{1}, \varphi_{2}, \phi_{1-}, \phi_{2-}),
  \label{phivec}
\end{align}
where the first and last two entries respectively represent the `upper' and `lower' 2D quantum Hall edges while the middle two represent the magnetic-domain-wall modes.  Physical charge-$e$ boson operators are correspondingly given by $a_{j\pm} = e^{-2i \phi_{j\pm}}$ and $b_j = e^{i \varphi_j}$ while the total charge density reads $\rho = \vec q\cdot \partial_x \vec\phi/\pi$.  

We can now define the non-chiral linear combinations
\begin{align}
& \theta_c =  \sum_{j=1,2} \left(\phi_{j+}+ \phi_{j-} + \varphi_{j}\right) \nonumber \\
& \theta_f =  \sum_{j=1,2} (-1)^j\left(\phi_{j+}+ \phi_{j-} + \varphi_{j}\right) \label{thetas}\\
& \theta_n =   \phi_{2-} - \phi_{2+}, \nonumber
\end{align}
which exhibit trivial commutation relations with one another.  
In addition the Hamiltonian terms
\begin{align}
H_\text{gap}= - \sum_{\alpha = c,f,n}\lambda_\alpha \cos(2 \theta_\alpha)
\end{align}
all arise from physical, charge-conserving bosonic processes and thus are allowed.  Assuming each piece is relevant the $\theta_{\alpha}$ fields lock to the minima of the respective cosine potentials thereby gapping each sector.  Repeating at all domain walls yields a fully gapped surface.

In essence, we have now seamlessly sewn together, in a $\T$-preserving way, $K = 2\sigma^x$ and $K = -2\sigma^x$ topological orders with the aid of the magnetic domain wall bosons.  Both topological orders support the same nontrivial quasiparticles: two flavors of charge $e/2$ bosons (`electric' and `magnetic') that bind to form a charge-$e$ fermion.  All three exhibit semionic mutual statistics with one another.  These excitations quite naturally survive as deconfined quasiparticles for our surface, which we can verify following Sec.~\ref{Statistics}.  

To construct strings that create well-separated $e/2$ dipoles, we first define operators that hop $e/2$ excitations across a single domain:
\begin{align}
&\mathcal{O}_{j,y+1/2}=\xi^\dagger_{j-,y+1} \xi_{j+,y} 
\end{align}
where $j = 1,2$ correspond to `electric' and `magnetic' dipoles.  The operators on the right side are given by
\begin{equation}
  \xi_{j-,y}=e^{- i \phi_{j-,y}}
\end{equation}
for either $j = 1$ or $2$ while 
\begin{align}
 &\xi_{1+,y}=e^{- i \phi_{1+,y}}b_{1,y}^\dagger b_{2,y}e^{- i (\phi_{1+,y}- \phi_{2+,y})}\\
 &\xi_{2+,y}=e^{- i \phi_{2+,y}}.
\end{align}
[Here and below we use a straightforward extension of Eqs.~\eqref{phivec} and \eqref{thetas} to arbitrary domain walls $y$.]  While somewhat cryptic, the nontrivial form of $\xi_{1+,y}$ above enables us to express extended vertical $e/2$ dipole operators in an appealing way,
\begin{align}
\hat v_i (y_1,y_2)&=\mathcal{O}_{i,y_2-1/2} \mathcal{O}_{i,y_2-3/2}\cdots \mathcal{O}_{i,y_1+3/2}\mathcal{O}_{i,y_1+1/2} \nonumber \\
&=\xi^\dagger_{i-,y_2}\sigma_i(y_1,y_2) \xi_{i+,y_1}
\end{align}
where
\begin{align}
 &\sigma_1(y_1,y_2)= e^{-i \sum_{y_1<y<y_2} \left(\theta_{n,y}-\theta_{f,y}\right) }\\
 &\sigma_2(y_1,y_2)= e^{-i \sum_{y_1<y<y_2} \theta_{n,y}}. 
\end{align}
In the gapped surface phase the intervening $\sigma_{1,2}$ strings above clearly obtain an expectation value and hence are invisible, so that $e/2$ quasiparticles are indeed deconfined.  

We can use the endpoint operators $\xi_{i\pm,y}^\dagger$---which create $e/2$ excitations with spin 1 as expected---to construct extended horizontal dipole operators as well,
\begin{align}
  \hat h_{i\pm,y} &=\xi^\dagger_{i\pm,y}(-\infty)\xi_{i\pm,y}(\infty) \nonumber.
\end{align}
As in Sec.~\ref{Statistics} we can now readily extract the braiding properties for the electric and magnetic $e/2$ quasiparticles [recall Eq.~\eqref{fullbraid}].  Since $[\xi_{i\pm,y},\sigma_i(-\infty,\infty)]=0$ both $e/2$ flavors exhibit trivial self-statistics.  Their mutual statistics is, however, nontrivial---a full braid of the electric particle around the magnetic particle yields a phase of $\pi$.  That is, they are mutual semions consistent with the intuition outlined above.  The expected properties of the charge-$e$ fermion follow by fusing the $e/2$ types.  We conclude that the surface hosts a $\T$-invariant cousin of 2D topological order with $K=2 \sigma^x$ in agreement with Refs.~\onlinecite{AshvinSenthil,ScottLesik}.

\subsection{$E_8$ bosonic topological insulator}

Vishwanath and Senthil~\cite{AshvinSenthil} further introduced a bosonic topological phase with time-reversal symmetry in which surface magnetic domain walls are identical to the edge of Kitaev's two-dimensional $E_8$ state\cite{ToricCode} (see also Refs.~\onlinecite{LuWires,Cano}).  Each domain wall binds eight non-fractionalized chiral boson modes, all of which co-propagate.  This set of modes is described by an eight-component vector of fields $\vec\phi$ that satisfies commutation relations
\beq
[\phi^I(x),\phi^J(x')] = i \pi (K_{E_8}^{-1})^{IJ} \sgn(x-x'),
\eeq
where $K_{E_8}$ is the Cartan matrix of the exceptional Lie group $E_8$.  (For an explicit form of $K_{E_8}$ see Refs.~\onlinecite{AshvinSenthil,LuWires,Cano}.)

Following the previous subsection we will construct a $\T$-invariant array of magnetically gapped domains, augmented by 2D phases that allow all surface modes to gap out via intra-domain-wall perturbations.  References \onlinecite{AshvinSenthil,BurnellChen} supply useful guidance for choosing these auxiliary 2D systems: they should comprise `half' of the $E_8$ state.  More precisely, the edge of the auxiliary phases should harbor four chiral modes, two sets of which can completely annihilate the $E_8$ magnetic domain-wall modes.  As we will see alternately tiling the surface with topologically ordered states described by the K-matrix
\begin{equation}
  K_{D_4} = \begin{pmatrix} 2 & -1 & -1 & -1 \\ -1 & 2 & 0 & 0 \\ -1 & 0 & 2 & 0 \\ -1 & 0 & 0 & 2 \end{pmatrix}.
  \label{KD4}
\end{equation}
in `even' domains and the time-reversed conjugate $(-K_{D_4})$ in `odd' domains does the job.  Topological orders with either $K_{D_4}$ or $-K_{D_4}$ share the same quasiparticle types---there are three nontrivial species with fermionic self-statistics, all of which exhibit mutual semionic statistics with one another.  One can view this structure as an all-fermion version of the topological order described in the last subsection.\cite{BurnellChen}  Furthermore, their edges support four copropagating modes as desired.  It is thus rather natural that one can seamlessly sew the $\pm K_{D_4}$ states together with the aid of the $E_8$ domain-wall degrees of freedom, yielding a $\T$-invariant fully gapped surface with the same three nontrivial quasiparticles in agreement with previous works\cite{AshvinSenthil,BurnellChen}.  [Note that these references refer to the topological order as SO(8)$_1$.  We denote the K-matrix with a subscript $D_4$ for reasons that will become clear below.]

Verifying that the decorated surface can be fully gapped out as claimed is in principle straightforward---one needs to identify eight commuting, non-chiral fields that can be `pinned' by physical domain-wall perturbations.  The numerous modes involved, however, render the usual K-matrix description a bit unwieldy.  Instead we will use the equivalent language of integer lattices.  Since the formalism may be less familiar to some readers we first provide a lightning review.  For additional background see Ref.~\onlinecite{Cano}, for example.

Consider a system described by an $N$-dimensional K-matrix $K$ together with a set of our usual $2\pi$-periodic fields $\phi^{1,\ldots,N}$ familiar from previous sections.  One can decompose the K-matrix in terms of primitive vectors $\mathbf{a}_{1,\ldots,N}$ that span an $N$-dimensional lattice via
\beq
  K_{IJ} = \mathbf{a}_I \circ \mathbf{a}_J \equiv \sum_{\mu = 1}^N s_\mu a^\mu_I a^\mu_J,
  \label{latticeK}
\eeq
with $s_\mu \in \{ +1,-1 \}$ chosen to reproduce the signature of $K_{IJ}$.  In this representation it is convenient to introduce chiral boson fields $\vec \chi$ that exhibit $2\pi$ periodicity under primitive-lattice-vector translations by defining
\beq
  \bm{\chi} \equiv \phi^I \mathbf{a}_I.
\eeq
(Here and below repeated indices are summed.)  Periodicity of our original $\phi_I$ fields indeed implies that
\beq
\bm{\chi} + 2\pi n^I \mathbf{a}_I \equiv \bm{\chi}, \quad n^I \in \mathbb{Z}.
\eeq

The dual lattice is spanned by primitive vectors $\mathbf{b}^{I} = (K^{-1})^{IJ} \mathbf{a}_J$; one can verify using Eq.~\eqref{latticeK} that $\mathbf{b}^I \circ \mathbf{a}_J = \delta^I_J$ as required.  Note that the inverse K-matrix takes an appealing form using dual lattice vectors that nicely parallels Eq.~\eqref{latticeK}:
\begin{equation}
  (K^{-1})^{IJ} = \mathbf{b}^I \circ \mathbf{b}^J.
\end{equation}
The dual lattice is also useful for constructing physical quasiparticle operators, which in our previous formulation take the form $e^{i l_I \phi^I}$ for integer $l_I$.  When expressed in terms of our $\bm{\chi}$ fields we obtain
\begin{equation}
  e^{i l_I \phi^I } = e^{i l_I \mathbf{b}^I \circ \bm{\chi}}.
  \label{PhysicalOps}
\end{equation}
Commutators of the latter follow from
\begin{align}
[l_I \mathbf{b}^I \circ \bm{\chi}(x), m_J \mathbf{b}^J \circ \bm{\chi}(x') ] &= i \pi (l_I \mathbf{b}^I) \circ (m_J \mathbf{b}^J)\, \sgn(x-x') \notag\\
&= i \pi l_I (K^{-1})^{IJ} m_J \sgn(x-x').
\end{align}

Local quasiparticle operators representing elementary bosons or fermions correspond to integer vectors $l_I = K_{IJ} n^J$ in Eq.~\eqref{PhysicalOps}, with $n^J \in \mathbb{Z}$.  For this special case we have $l_I \mathbf{b}^I = n^I \mathbf{a}_I$---i.e., the vector multiplying $\bm{\chi}$ in Eq.~\eqref{PhysicalOps} is an element of both the dual \emph{and} direct lattices.  We will be particularly interested in sets of mutually commuting, non-chiral fields that can be simultaneously pinned by Hamiltonian terms built from products of such local operators.  In the language of lattices, such sets correspond to mutually orthogonal null vectors in the direct lattice [using the inner product defined in Eq.~\eqref{latticeK}].  This geometric interpretation is the main virtue of the alternative formulation reviewed here and greatly facilitates our forthcoming analysis.

Let us now apply this formalism to our decorated surface.  Each domain wall hosts eight co-propagating $K_{E_8}$ modes together with two sets of backward-propagating modes corresponding to $K_{D_4}$ in Eq.~\eqref{KD4}.  The K-matrix describing this collection clearly breaks up into three decoupled blocks, each of which corresponds to its own lattice.  In light of the chiralities we can take the signs $s_\mu$ in Eq.~\eqref{latticeK} to be $+1$ for the $K_{E_8}$ block and $-1$ for the $K_{D_4}$ blocks (or vice versa).  The former corresponds to the eight-dimensional $E_8$ lattice whose structure admits a simple description in Cartesian coordinates: the components of lattice vectors should be either all integers or all half-integers, and should sum to an even number.  Examples of primitive vectors are $(1,1,0,0,0,0,0,0)$ and $(1,1,1,1,1,1,1,1)/2$, which both have norm 2 under the inner product defined above.  Each of the remaining blocks corresponds to the four-dimensional $D_4$ lattice.  The Cartesian coordinates of the $D_4$ lattice are all integers that sum to an even number---e.g., $(1,1,0,0)$ and $(1,-1,0,0)$---which have negative norm because of the different signature relative to the $E_8$ block.  It follows from these descriptions that $E_8$ represents two copies of $D_4 \times D_4$ shifted by $(1,1,1,1,1,1,1,1)/2$ relative to each other.

Armed with this geometric language, we can readily construct the perturbations necessary to gap all fields at a domain wall.  The required interactions all take the form 
$\cos(\mathbf{v}_8 \circ\bm{\chi}_8 + \mathbf{v}_{4\times4}\circ\bm{\chi}_{4\times4})$, where $\mathbf{v}_8$ is in $E_8$ and $\mathbf{v}_{4\times4}$ is in $D_4 \times D_4$ while $\bm{\chi}_8$ and $\bm{\chi}_{4\times 4}$ represent the corresponding fields.  Because the lattice structures match up so nicely, choosing $\mathbf{v}_{4\times4} = \mathbf{v}_8 \equiv \mathbf{v}$ guarantees that the argument of the cosine yields a nonchiral field that can be pinned by the interaction.  We can further construct a set of simultaneously pin-able nonchiral fields sufficient to completely gap the domain wall simply by identifying eight orthogonal vectors $\{\mathbf{v}\}$.  One such choice reads
\begin{align}
&(1,1,0,0,0,0,0,0), \; (1,-1,0,0,0,0,0,0), \notag\\
&(0,0,1,1,0,0,0,0), \; (0,0,1,-1,0,0,0,0), \notag\\
&(0,0,0,0,1,1,0,0), \; (0,0,0,0,1,-1,0,0), \notag\\
&(0,0,0,0,0,0,1,1), \; (0,0,0,0,0,0,1,-1).
\end{align}
It is worth noting that the ``interaction'' corresponding to each vector above just describes tunneling of single bosons between the $E_8$ edge and one of the $D_4$ edges.  Moreover, because these are all primitive lattice vectors such perturbations gap the domain walls trivially without generating further fractionalization.  The auxiliary $\pm K_{D_4}$ topological orders thus indeed can seamlessly stitch together at the bosonic TI surface, as claimed above.

\section{Discussion}
\label{Discussion}

As the theoretical understanding of free-fermion symmetry-protected topological phases enters maturity, increasing attention has naturally turned to exploring nontrivial effects of strong interactions on such states.  A rich story is now emerging.\cite{SenthilReview}  Previous work has demonstrated that interactions allow one to $(i)$ smoothly connect certain phases that are topologically distinct in the free-particle limit,\cite{FidkowskiKitaev1,Gurarie,RyuZhang,QiInteractions,YaoRyu,GuLevin,Neupert,FidkowskiTSC,WangSenthil,MetlitskiTO2} $(ii)$ generate new short-range-entangled topological phases with no non-interacting counterpart, for both bosonic and fermionic systems,\cite{AshvinSenthil,WangClassification,Clustons} and $(iii)$ form gapped topological orders at the boundary of a 3D topological phase that can not exist in strictly 2D systems with the same symmetry\cite{AshvinSenthil,BondersonTO,WangTO,ChenTO,MetlitskiTO,BurnellChen,ScottLesik,FidkowskiTSC,WangSenthil,ChenBurnell,MetlitskiTO2}.  To this list we have added a new possibility: strong interactions can nucleate exotic \emph{gapless} composite Dirac liquids at the surface of a 3D electronic TI.  

The hallmark of these states is the existence of a single Dirac cone formed not of electrons, but rather emergent neutral fermions.  We constructed a hierarchy of such states.  The `fundamental' composite Dirac liquid arises by effectively stripping the charge off of the original surface electrons, yielding a first-generation neutral Dirac cone.  A nested composite Dirac liquid emerges upon similarly stripping a `pseudocharge' off of these neutral fermions, resulting in a second-generation neutral Dirac cone at the surface (and so on).  All of these states display unusual transport characteristics.  In charge measurements composite Dirac liquids appear rather featureless---they are incompressible, electrical insulators.  Because of the neutral Dirac cone, however, they exhibit `metallic' heat conduction similar to the surface state of a noninteracting electronic TI.  The gapless nature of composite Dirac liquids allowed us to access nontrivial descendant phases by gapping the neutral Dirac cone via pairing---which preserves electron number conservation---or magnetism.  Quite remarkably, pairing the first- and second-generation neutral Dirac cones respectively produces the symmetric T-Pfaffian and Pfaffian-antisemion phases captured by completely different means in previous studies\cite{BondersonTO,WangTO,ChenTO,MetlitskiTO}.  The perspective developed here thus establishes a hierarchical relationship between these topological orders.  Magnetically gapping the neutral Dirac cones instead yields Abelian topological orders.  We showed that all of these descendants can be mapped to weak- and strong-pairing phases of 2D composite Fermi liquids upon breaking symmetries.   

On a technical level, demonstrating that a particular strongly correlated phase arises from a given Hamiltonian poses a notoriously difficult problem.  Here we overcame this challenge by exploiting a quasi-1D deformation of the electronic TI surface that allowed us to controllably access composite Dirac liquids and descendants thereof starting from an electronic Hamiltonian.  The price we paid is abandonment of local time-reversal symmetry $\mathcal{T}$ in favor of a weaker `antiferromagnetic' operation $\T$ (corresponding to $\mathcal{T}$ composed with a translation).  It would be extremely interesting in future work to develop a theory for composite Dirac liquids that restores the local time-reversal symmetry in which we are ultimately interested.  A $\mathcal{T}$-invariant formulation would allow one to address subtle issues such as whether time reversal squares to $+1$ or $-1$ when acting on the non-Abelian excitations in the T-Pfaffian\cite{ChenTO}, which we do not know how to treat in our setup.  Finding spatially  isotropic, $\mathcal{T}$-invariant Hamiltonians supporting either composite Dirac liquids or their descendants as ground states remains another important open question.  Our work does, however, provide some possible guidance in this direction.  For instance, our `intrinsic' composite Dirac liquid construction (Sec.~\ref{intrinsic}) hints that multiple Dirac cones may, counterintuitively, aid in the stabilization of exotic surface phases.  In particular, studying an isotropic analogue of the charge-gapping terms provided in Eq.~\eqref{ChargeGapIntrinsic} for a surface with three electronic Dirac cones could prove illuminating. 

We adapted similar quasi-1D surface deformations that relax $\mathcal{T}\rightarrow \T$ symmetry to controllably study nontrivial phases at the boundary of two different types of \emph{bosonic} 3D TIs.  In both cases we recovered topological orders in agreement with previous works that treated isotropic systems with local time-reversal symmetry\cite{AshvinSenthil,BurnellChen,ScottLesik}.  Given these successes it is natural to ask when a symmetric surface can be reliably distilled to a quasi-1D setup with reduced symmetry, while preserving the structure of possible surface phases.  We conjecture that this is always possible \emph{provided} reducing the symmetry in the same way throughout the system does not destroy the topological properties of the bulk.  As a concrete example, the quantized magnetoelectric effect for a 3D electronic TI with local time-reversal symmetry remains upon introducing antiferromagnetic order in the bulk,\cite{AFTImong} which is ultimately why our construction works for this case.  However, the preservation of bulk topological physics on lowering symmetry is certainly not guaranteed.  

Three-dimensional topological superconductors constitute an interesting example where relaxing $\mathcal{T}$ to $\T$ is not always benign.  Without interactions such phases are characterized by an integer invariant $\nu\in \mathbb{Z}$ protected by local time-reversal symmetry $\mathcal{T}$.\cite{Schnyder3d,TRItopologicalSC3}  A $\mathcal{T}$-invariant surface correspondingly hosts $\nu$ gapless Majorana cones.  Upon introducing strong interactions previous work shows that one can gap out these Majorana cones without breaking any symmetries.\cite{FidkowskiTSC,WangSenthil,MetlitskiTO2}  The symmetric gapped surfaces necessarily display topological order except when $\nu$ is a multiple of 16 (in that case the surfaces are trivial).  This result implies that the integer free-fermion classification of topological superconductors reduces to $\mathbb{Z}_{16}$ with interactions.\cite{FidkowskiTSC,WangSenthil,MetlitskiTO2}  

Following our treatment of electronic TIs, one might hope to recover these results using a quasi-1D deformation of the surface obtained from a $\T$-invariant array of magnetic domain walls.  For a non-interacting topological superconductor with integer invariant $\nu$, each domain wall can support $\nu$ copropagating chiral Majorana modes.  For \emph{any} even $\nu$, however, all of these modes can trivially gap out without breaking $\T$ symmetry by democratically annihilating pairs of  counterpropagating modes from adjacent domain walls [similar to Fig.~\ref{EdgeFig}(a)].  The physics is clearly then rather different from the case with local time-reversal symmetry.  The distinction here reflects the fact that introducing antiferromagnetic order into the bulk of a noninteracting 3D topological superconductor reduces the integer classification down to $\mathbb{Z}_2$ (in sharp contrast to the situation for topological insulators).  See Appendix \ref{AFTSC} for a derivation of the classification.  These observations support our conjecture above.  (Note that this classification agrees with that given in Ref.~\onlinecite{WangTSC} based on macroscopic thermal responses, which nicely parallels the relation between the TI and the antiferromagnetic TI.\cite{AFTImong})  We expect that it is still possible to faithfully treat the full set of strongly interacting 3D topological superconductors from a quasi-1D construction.  To do so one would need to invoke additional symmetries that enlarge the $\mathbb{Z}_2$ classification for antiferromagnetic topological superconductors back to $\mathbb{Z}$.  Pursuing such methods to capture exotic parent gapless states and descendant topological orders in such settings---and other symmetry protected topological phases---poses an interesting avenue for future research.

\acknowledgments{We are indebted to David Clarke, Lukasz Fidkowski, Matthew Fisher, Eduardo Fradkin, Roger Mong, Lesik Motrunich, Yuval Oreg, Xiaoliang Qi, T.~Senthil, Ari Turner, and Ashvin Vishwanath for illuminating conversations on this work.
We also acknowledge funding from the NSF through grant DMR-1341822 (J.\ A.); the Alfred P.\ Sloan Foundation (J.\ A.); the Caltech Institute for Quantum Information and Matter, an NSF Physics Frontiers Center with support of the Gordon and Betty Moore Foundation through Grant GBMF1250; and the Walter Burke Institute for Theoretical Physics at Caltech.
}

\appendix

\section{Majorana zero modes in the T-Pfaffian phase}
\label{MajoranaAppendix}

Section \ref{TPfaffian} argued that in the T-Pfaffian phase charge-$e/4$ quasiparticles---which appear as $h/(2e)$ flux seen by the paired neutral fermions---bind Majorana zero modes.  In this Appendix we back up this assertion by explicitly examining the Hamiltonian with an isolated $e/4$ excitation.  Consider the fermionic sector of the theory that describes the pairing-gapped neutral Dirac cone:
\begin{eqnarray}
  H_n +H_{\rm pair} &=& \sum_y (-1)^y \int_x \bigg{\{} +i \hslash \tilde v_x \tilde \psi_y^\dagger \d_x \tilde \psi_y 
  \nonumber \\
  &-& \tilde t \left[\tilde \psi_y^\dag \tilde \psi_{y+1}e^{2i(\Theta_{c,y}+\Theta_{c,y+1})} + \mathrm{H.c.}\right] 
  \nonumber \\
  &+&  \left[\tilde \Delta \tilde \psi_y \tilde \psi_{y+1}e^{-2i(\Theta_{c,y}-\Theta_{c,y+1})} + \mathrm{H.c.}\right]\bigg{\}}.
  \label{MajoranaH}
\end{eqnarray}
Note that it is now convenient to work with `undressed' neutral fermions $\tilde \psi_y$ so that the dependence on the charge field $\Theta_{c,y}$ is explicit.  

For simplicity we specialize to the limit $\tilde t = \tilde \Delta$ and place an $e/4$ excitation at $x = 0$ in an even chain $y_0 \in 2\mathbb{Z}$.  We can then take the corresponding charge-field configuration to be 
\begin{equation}
  \Theta_{c,y_0}(x) = \begin{cases}
    0, & x <0\\
    \pi/ 4,              & x > 0
\end{cases}
\end{equation}
with $\Theta_{c,y\neq y_0} = 0$.  By examining Eq.~\eqref{MajoranaH} one sees that the $e/4$ kink modifies only pairing and tunneling terms with $y = y_0$ and $y_0\pm 1$.  After some algebra one can group these terms as follows,
\begin{eqnarray}
  &&- \tilde t \int_x e^{2i\Theta_{c,y_0}(x)}[(\tilde \psi_{y_0}^\dagger + \sgn(x)\tilde \psi_{y_0})(\tilde \psi_{y_0+1}^\dagger + \tilde \psi_{y_0+1})
  \nonumber \\
  &&+ (\tilde \psi_{y_0-1}^\dagger - \tilde \psi_{y_0-1})(\sgn(x)\tilde \psi_{y_0}^\dagger -\tilde \psi_{y_0})]
\end{eqnarray}
It is illuminating to now decompose the neutral fermions in terms of Majorana operators via $\tilde \psi_y = \eta_{1,y} + i \eta_{2,y}$.  In this representation the $e/4$ excitation alters the connections between Majoranas in a simple way; for instance, $\eta_{2,y_0}$ couples to $\eta_{1,y_0+1}$ at $x<0$ but $\eta_{2,y_0-1}$ at $x > 0$.  The Hamiltonian for the four Majorana operators involved (including kinetic terms) reads
\begin{eqnarray}
  H_{\rm Maj} &=& \int_x[i \hslash \tilde v_x(\eta_{1,y_0}\partial_x \eta_{1,y_0} + \eta_{2,y_0}\partial_x \eta_{2,y_0} 
  \nonumber \\
  &-& \eta_{1,y_0+1}\partial_x \eta_{1,y_0+1} - \eta_{2,y_0-1}\partial_x \eta_{2,y_0-1})
  \nonumber \\
  &+& 4i \tilde t\Theta(-x)(\eta_{2,y_0}\eta_{1,y_0+1} -\eta_{2,y_0-1}\eta_{1,y_0})
  \nonumber \\
  &-& 4i \tilde t\Theta(x)(\eta_{1,y_0}\eta_{1,y_0+1} -\eta_{2,y_0-1}\eta_{2,y_0})]
\end{eqnarray}
with $\Theta(x)$ the Heaviside step function.  We stress that this set of Majorana operators decouples from all others at the $\tilde t = \tilde \Delta$ point studied here.  

The existence of a Majorana zero mode bound to the $e/4$ charge becomes manifest upon rewriting $H_{\rm Maj}$ in a convenient basis.  As a first step we define new Majorana operators
\begin{equation}
  \eta_{\pm} = \frac{\eta_{1,y_0} \pm \eta_{2,y_0}}{\sqrt{2}},~~~~~~ \eta'_\pm = \frac{\eta_{1,y_0+1} \pm \eta_{2,y_0-1}}{\sqrt{2}},
\end{equation}
which allows us to write
\begin{eqnarray}
  H_{\rm Maj} &=& \int_x[i \hslash \tilde v_x \sum_{s = \pm}(\eta_{s}\partial_x \eta_{s} - \eta'_{s}\partial_x \eta'_{s} )
  \nonumber \\
 &-&4i\tilde t[\eta_-\eta_-' + \sgn(x) \eta_+ \eta'_+].
\end{eqnarray}
The counterpropagating $\eta_-$ and $\eta_-'$ Majoranas gap out trivially due to the first $\tilde t$ coupling on the second line.  There is, however, a sign-changing domain wall in the hybridization term for $\eta_+$ and $\eta_+'$ that binds the zero mode.  In fact, upon defining new complex fermions $\psi_L = \eta_+ + i \eta_-$ and $\psi_R = \eta_-' + i \eta_+'$ the problem maps precisely to a quantum spin Hall edge gapped by magnetism on one end and superconductivity on the other:
\begin{eqnarray}
  H_{\rm Maj} &=& \int_x[-i \hslash \tilde v_x (\psi_R^\dagger \partial_x \psi_R - \psi_L^\dagger \partial_x \psi_L)
  \nonumber \\
 &+&2\tilde t[\Theta(x)\psi_R \psi_L + \Theta(-x)\psi_R^\dagger \psi_L + H.c.].
\end{eqnarray}
A domain wall separating these two types of incompatibly gapped regions is known to harbor a protected Majorana zero mode,\cite{MajoranaQSHedge} and hence so does the $e/4$ quasiparticle in the T-Pfaffian phase.

\section{Classification of antiferromagnetic topological superconductors}
\label{AFTSC}

Consider a clean, noninteracting 3D topological superconductor preserving time-reversal symmetry $\mathcal{T}$ such that $\mathcal{T}^2 = -1$.  Such states are characterized by an integer invariant $\nu$.\cite{Schnyder3d,TRItopologicalSC3} Here we show that upon introducing antiferromagnetic order in the bulk, the relevant topological invariant becomes $\nu$ mod 2---i.e., among the infinite set we started with only two topologically distinct phases remain.  Closely following Mong et al.,\cite{AFTImong} we add an antiferromagnetic order parameter to the topological superconductor that doubles the unit cell and hence modifies the classification by topological invariants defined over the Brillouin zone.  Despite the violation of $\mathcal{T}$ the system nevertheless retains a quantized thermal response\cite{RyuMooreLudwig} protected by a remnant antiunitary symmetry $\tilde{\mathcal{T}} = \mathcal{T} T_{1/2}$, where $T_{1/2}$ corresponds to translation through half of the new unit cell.  For concreteness suppose that $T_{1/2}$ corresponds to translation along the $z$ direction.  

Doubling the unit cell folds the Brillouin zone such that the plane $(k_x,k_y,\pi)$ of the undoubled system maps to $(k_x,k_y,0)$.  The symmetry $\tilde{\mathcal{T}}$ squares to $-1$ when acting on states with momenta that reside in this plane, so the system admits a characterization by a \emph{two-dimensional} topological invariant.  One can compute this invariant explicitly in the limit where the order parameter is weak enough that the system adiabatically connects to a superconductor with full time-reversal symmetry.  Schnyder et al.\cite{RyuTenfoldWay} (whose argument for topological superconductors closely follows that of Qi et al.\cite{TopologicalFieldTheoryTI} for topological insulators) show that the two-dimensional invariant takes values in $\mathbb{Z}_2$---that is, there is only one nontrivial kind.   Whether the invariant is trivial or nontrivial can be determined by smoothly extending the two-dimensional `band structure' into three dimensions: it is nontrivial if this extension has an odd two-dimensional topological invariant and trivial if the extension has an even invariant.  
Turning the argument around, the antiferromagnetic topological superconductor is nontrivial if it smoothly connects to a $\mathcal{T}$-invariant 3D topological superconductor with odd invariant $\nu$, and trivial otherwise.  

\bibliography{SurfaceTO}

\end{document}